\documentclass[fleqn,usenatbib]{mnras}

\usepackage{newtxtext,newtxmath}

\usepackage[T1]{fontenc}
\usepackage{float}

\DeclareRobustCommand{\VAN}[3]{#2}
\let\VANthebibliography\thebibliography
\def\thebibliography{\DeclareRobustCommand{\VAN}[3]{##3}\VANthebibliography}

\usepackage{graphicx}	
\usepackage{amsmath}	
\usepackage{multirow}
\usepackage[export]{adjustbox}
\usepackage{subfig}
\usepackage{orcidlink}
\usepackage{threeparttable}

\newcommand{\hi}{H{\sc i}}
\newcommand{\FCNM}{$f_\mathrm{CNM}$}
\newcommand{\kms}{km s$^{-1}$}

\defcitealias{marchal_2024}{M24}
\defcitealias{heiles_millennium_2003-1}{HT03}

\title[Cold Skeleton of the Mag. Clouds and Bridge]{Revealing the cold skeleton of the Magellanic Clouds and the Magellanic Bridge with ASKAP}

\author[J. Dempsey et al.]{James Dempsey\,\orcidlink{0000-0002-4899-4169}$^{1,2}$\thanks{E-mail: james.dempsey@anu.edu.au},
N.~M. McClure-Griffiths\,\orcidlink{0000-0003-2730-957X}$^{1}$,
Antoine Marchal\,\orcidlink{0000-0002-5501-232X}$^{1,3,4}$,
S. E. Clark\,\orcidlink{0000-0002-7633-3376}$^{5,6}$, \newauthor 
John M. Dickey\,\orcidlink{0000-0002-6300-7459}$^{7}$,
Min-Young Lee\,\orcidlink{0000-0002-9888-0784}$^{8}$,
Claire Murray\,\orcidlink{0000-0002-7743-8129}$^{9,10}$,
Hiep Nguyen\,\orcidlink{0000-0002-2712-4156}$^{1}$,
Nickolas M. Pingel\,\orcidlink{0000-0001-9504-7386}$^{11}$,\newauthor 
Snežana Stanimirović\,\orcidlink{0000-0002-3418-7817}$^{12}$,
Jacco Th. van Loon\,\orcidlink{0000-0002-1272-3017}$^{13}$ 
Helga Dénes\,\orcidlink{0000-0002-9214-8613}$^{14}$,
Steven J. Gibson\,\orcidlink{0000-0002-1495-760X}$^{15}$,  \newauthor 
Katie Jameson\,\orcidlink{0000-0001-7105-0994}$^{16}$,  
Ian Kemp\,\orcidlink{0000-0002-6637-9987}$^{17}$,  
Callum Lynn\,\orcidlink{0000-0001-6846-5347}$^{1}$,
and Yik Ki Ma\,\orcidlink{0000-0003-0742-2006}$^{1,18}$,
\\
\\
$^{1}$Research School of Astronomy and Astrophysics, The Australian National University, Canberra, ACT 2611, Australia\\
$^{2}$CSIRO Information Management and Technology, GPO Box 1700 Canberra, ACT 2601, Australia\\
$^{3}$ Laboratoire de Physique de l’École Normale Supérieure, ENS, Université PSL, CNRS, Sorbonne Université, Université Paris Cité, F75005, Paris, France
\\
$^{4}$ LUX, Observatoire de Paris, Université PSL, Sorbonne Université, 75014 Paris, France
\\
$^{5}$Department of Physics, Stanford University, Stanford, CA 94305, USA\\
$^{6}$Kavli Institute for Particle Astrophysics \& Cosmology, P.O. Box 2450, Stanford University, Stanford, CA 94305, USA\\
$^{7}$School of Natural Sciences, University of Tasmania, Hobart, TAS 7005, Australia\\
$^{8}$Korea Astronomy and Space Science Institute, 776, Daedeokdae-ro, Yuseong-gu Daejeon 34055, Republic of Korea\\
$^{9}$Department of Physics \& Astronomy, Johns Hopkins University, 3400 N. Charles Street, Baltimore, MD 21218, USA\\
$^{10}$Space Telescope Science Institute, 3700 San Martin Drive, Baltimore, MD 21218, USA\\
$^{11}$Department of Astronomy, Indiana University, 727 East Third Street, Bloomington, IN 47405, USA\\
$^{12}$Department of Astronomy, University of Wisconsin--Madison, 475 N Charter St., Madison, WI 53706, USA \\
$^{13}$Lennard-Jones Laboratories, Keele University, ST5 5BG, UK
$^{14}$College of Sciences and Engineering, Universidad San Francisco de Quito, Quito, Ecuador \\
$^{15}$Department of Physics and Astronomy, Western Kentucky University, Bowling Green, KY 42101, USA\\
$^{16}$Caltech Owens Valley Radio Observatory, Pasadena, CA 91125, USA\\
$^{17}$International Centre for Radio Astronomy Research (ICRAR), Curtin University, Bentley, WA 6102, Australia\\
$^{18}$Max-Planck-Institut f\"ur Radioastronomie, Auf dem H\"ugel 69, 53121 Bonn, Germany \\ 
}

\date{Accepted 2026 May 05. Received 2026 May 05; in original form 2026 February 03}

\pubyear{2026}

\begin{document}
\label{firstpage}
\pagerange{\pageref{firstpage}--\pageref{lastpage}}
\maketitle

\begin{abstract}
We present the GASKAP-HI pilot absorption survey of neutral hydrogen (\hi) in the Magellanic system.
This survey provides 3219 sightlines across the Large (LMC) and Small Magellanic Clouds (SMC) and the Magellanic Bridge (MB) towards 1.4-GHz continuum sources, representing a 15-fold increase on pre--GASKAP-HI sampling of the Magellanic System. 
We find 344 candidate detections of cold gas at Magellanic velocities (v$_\text{LSRK} \geq 90$ km s$^{-1}$), with signal-to-noise ratio $> 3$ detection rates of 
44\% (LMC; 192 of 438),
73\% (SMC; 85 of 117) and 
4\% (MB; 35 of 793). 
We examine the candidate detections within the MB, Gaussian decompose these and examine the cold gas across the MB.
Here we find that the majority of cold gas detections are found closer to the SMC. 
We also find potential evidence of the recent formation of cold gas on the outskirts of a shell within the MB.
We find a mean cold gas fraction of $\overline{f_\text{CNM}} = 0.12 \pm 0.08$ for the MB, which is very similar to the SMC and lower than the LMC value of 0.14. 
Overall, we reveal cold gas distributed extensively across the Magellanic system, including within the MB, and surmise that the cold gas in the MB is either pulled from the SMC as part of the formation of the MB, or formed in the turbulence of those same interactions.
\end{abstract}


\section{Introduction}

Neutral atomic hydrogen (\hi) is the most abundant component of the interstellar medium (ISM).
It traces both large- and small-scale structures in galaxies.
\hi\ is found in two long-lasting phases, the diffuse warm neutral medium (WNM; $T = 4,000 - 8,000$ K\footnote{Temperatures ranges are typical for the Solar neighbourhood}) and the dense cold neutral medium (CNM; $T = 25 - 250$ K\footnotemark[\value{footnote}]) with an unstable, lukewarm phase (UNM; $T = 250 - 4,000$ K\footnotemark[\value{footnote}]) between them \citep{wolfire_2003,bialy_2019}.
The cooling of WNM to CNM and onwards to molecular clouds and star formation is driven by ISM dynamics such as gas flow collisions, and winds driven by supernovae and star formation. 
This cooling requires shielded clouds and filaments and is one of the defining processes of star formation \citep{mcclur23}.
Thus it is crucial to understand the extent and location of cold gas in order to explore the future of star formation in a region.

Both WNM and CNM produce 21-cm emission through the \hi\ spin-flip transition, contributing to the overall brightness temperature ($T_\text{b}$) that we observe. 
However, the weaker emission from the CNM can be difficult to disentangle from the stronger WNM emission.
Instead, we frequently turn to absorption, where the cold gas blocks continuum emission from a background source.
This allows us to directly measure the cold \hi\ gas. 
We can also then quantify the CNM characteristics by matching the absorption measurements with emission observations as close as possible in scale and proximity.
When the absorption measurements are combined with information about the emission in the same region, we can estimate the spin or excitation temperature ($T_s$) of the cold gas (\cite{heiles_millennium_2003-1}, HT03).
In the dense CNM regime, the spin-flip transition is thermalised due to collisions and so the kinetic temperature of the gas ($T_k$) can be approximated as $T_s \approx T_k$ \citep{field_thermal_1965,liszt_2001}.
Thus, we can measure the kinetic temperature of the gas along with other properties such as turbulent broadening, and mass fraction.

While the CNM is difficult to detect in emission, its effects are still present.
Decomposition of emission spectra into Gaussian components has been used to tease out the narrow spectral features indicative of the CNM \citep{haud_2000,nidever_2008,kalberla_haud_2015_gass,marchal_2019,murray_2020,lei_2025}.
These approaches provide both spatial and spectral information about the different gas phases, but can be computationally prohibitive for high resolution data covering large areas.
As a result, cold \hi\ gas has not yet been mapped in detail across the scale of an entire interacting system.
An alternative approach, to provide spatial but not spectral information, was demonstrated by \citet[][hereafter M24]{marchal_2024}.
In this technique, the Fourier transform (FT) of the emission spectrum ($T_b(v)$) is used to assess the cold gas mass fraction, thus highlighting the potential presence of cold gas.
The technique provides a lower limit on the fraction of \hi\ in the cold phase and allows that lower limit to be efficiently mapped across a large data set.
The advantage of this approach is that it gives us information about the location of the CNM even in the absence of background sources.
The disadvantage of the technique is that it does not provide a strict quantitative measure of CNM fraction.

Our closest major neighbours in the Local Group are the interacting pair of galaxies, the Large (LMC) and Small Magellanic Clouds (SMC) along with the debris fields between them -- the Magellanic Bridge (MB) -- and around them -- the Magellanic Stream (MS) and the Leading Arm (LA).
These galaxies give us a front row seat to the dynamics of merging galaxies and interaction with the halo of our own Milky Way (MW).
Over the past three decades, the CNM of the Magellanic system has been explored with targeted absorption surveys, first of the brightest sources in the region  \citep{mebold_1991,kobulnicky_1999,matthews_2009} and then surveys of selected sources in the LMC \citep{dickey_1994,marx-zimmer_2000,liu_2021} and the SMC \citep{dickey_2000,jameson_2019}.
In common for all of these surveys was that they required long observations on each target source, making each source expensive to observe. 
As a result, source selection often favoured targets in higher density regions, or brighter continuum sources, where the chance of detections was greater.

The Australian Square Kilometre Array Pathfinder (ASKAP; \citealt{hotan_2021}) telescope, in contrast, allows us to measure all sources in its $5^{\circ} \times 5^{\circ}$ field down to a sensitivity limit set by dwell time.
The Galactic ASKAP \hi\ (GASKAP-HI; \citealt{dickey_2013}) project is using this capability to provide unbiased surveys of multiple fields in both absorption and emission simultaneously. 
The effectiveness of this approach was shown in pilot phase I CNM surveys of the SMC  \citep{dempsey_2022} and the outer MW \citep{dickey_2022}.

In this paper we explore the unprecedented view of cold \hi\ gas in the LMC, SMC and MB that the GASKAP-HI pilot phase I and II observations provide.
We present the full data set of Magellanic \hi\ absorption data from these observations, and then explore the CNM of the Bridge region as revealed by these data.
Future GASKAP-HI collaboration papers will examine the CNM of the LMC \citep[e.g.][in preparation]{chen_2026}, SMC and surrounding regions using this data set.
In section \ref{sec:data} we present the \hi\ absorption measurements by GASKAP-HI of the Magellanic system, the counterpart of the Milky Way foreground spectra described by \cite{nguyen_2024}. 
In section \ref{sec:bridge-abs} we use the MB absorption as a case study for exploring the CNM in a region that has only been sparsely sampled in the past.
Finally, we summarise our findings in section \ref{sec:conclusion}.

\section{Data}
\label{sec:data}

The focus of observations for the GASKAP-HI project's pilot phase II was shallow coverage of the central Magellanic system. 
In contrast, the full GASKAP-HI survey will cover a much wider region, taking in much of the Galactic Plane ($270$\degr $\le l \le$ $17$\degr  
), the Galactic Centre, much of the near part of the Magellanic Stream, as well as the Magellanic Clouds. 
In addition, the full survey will be much deeper, with total dwell times ranging from 30 hours on Magellanic Stream fields up to 200 hours on the Magellanic Clouds.

In August 2021, three 7-hour GASKAP-HI observations of the SMC were taken and stacked.
Between October 2021 and April 2022 nine fields across the LMC and MB were observed for $\sim$10.2 hours each.
A list of the observations is shown in Table \ref{tab:observations}.
All observations were within the acceptable flagging tolerances of the GASKAP-HI project, resulting in excellent coverage of both long baselines for absorption sensitivity and short baselines for emission sensitivity.
The emission data were also combined with GASS \citep{kalberla_haud_2015_gass}, as described in \cite{pingel_2022}, to provide coverage of the large-scale structure.
The Milky Way foreground spectra from these pilot phase II observations are described in \cite{nguyen_2024}. 
To expand the spatial coverage and improve sensitivity at Magellanic velocities, these observations have been combined in this dataset with GASKAP-HI Pilot phase I observations of the SMC, MB and Stream. 
These data were taken between December 2019 and June 2020.
Again, the emission data were combined with GASS.
It should be noted that the pilot phase I fields were in a different orientation from the GASKAP-HI Pilot phase II observations, resulting in overlaps and thus differing sensitivity in the SMC and western Bridge regions as compared to the the rest of Magellanic system.
In particular, there are two small `holes' of very low sensitivity (primary beam power $< 25$\%) centred around $\alpha=03h18m00s$ $\delta=-74d50m00s$ (below the middle of the MB) and $\alpha=01h50m00s$ $\delta=-74d15m00s$ (where the MB meets the SMC wing).
The telescope configuration for both phases I and II was the same as described in \cite{dempsey_2022} and the emission data characteristics are show in Table \ref{tab:proc-params}.

\begin{table}
  \caption{List of GASKAP-HI Pilot observations showing the pilot phase, ASKAP scheduling block ID (SBID), field identifier, equivalent exposure duration after flagging and the pointing centre of each observation.}
  \label{tab:observations}\centering
\begin{tabular}{ c c c r r r }
\hline \hline
 Phase & SBID & Field & Duration (hours) & R.A. (hms) & Dec. (dms)  \\ 
 \hline
I & 10941 & SMC & 9.51 & 00:58:43 & $-$72.31.49 \\
I & 10944 & SMC & 9.54 & 00:58:43 & $-$72.31.49 \\
I & 14180 & MB & 10.19 & 02:31:11 & $-$74.39.16 \\
I & 14158 & MS & 10.15 & 01:44:55 & $-$70.32.31 \\
I & 14211 & MS & 10.17 & 01:44:55 & $-$70.32.31 \\
II & 30584 & SMC & 7.26 & 00:52:43 & $-$72.47.45 \\
II & 30625 & SMC & 7.01 & 00:52:43 & $-$72.47.45 \\
II & 30665 & SMC & 7.16 & 00:52:43 & $-$72.47.45 \\
II & 38373 & 1 & 10.16 & 04:44:07 & $-$64.33.05 \\
II & 38791 & 2 & 10.16 & 05:33:47 & $-$65.11.23 \\
II & 38814 & 3 & 10.15 & 04:32:25 & $-$69.04.38 \\
II & 33047 & 4 & 10.15 & 05:33:00 & $-$69.51.28 \\
II & 38758 & 5 & 10.16 & 04:14:39 & $-$73.31.55 \\
II & 38845 & 6 & 10.16 & 05:31:47 & $-$74.31.50 \\
II & 38509 & 7 & 10.17 & 03:44:29 & $-$77.49.48 \\
II & 38466 & 8 & 10.15 & 03:11:44 & $-$71.02.31 \\
II & 38215 & 9 & 10.17 & 01:29:45 & $-$77.33.19 \\
\hline
\end{tabular}
\end{table}

\begin{table}
  \caption{Emission data characteristics for the GASKAP-HI Pilot fields.}
\centering
\begin{tabular}{ p{30mm} p{45mm} }
\hline \hline
Parameter & Value (units) \\
\hline
Pixel scale & 7$^{\prime\prime}$x7$^{\prime\prime}$ \\
Niter & 25000 \\
Mgain & 0.7 (peak flux reduced by 70\% to trigger
next major cycle) \\
Multiscale scales & Auto (determined by WSClean) \\
Multiscale-scale bias & 0.85 (bias towards smaller scales) \\
Robust parameter & 0.75 \\
Taper (applied in addition to robust
weighting) & 14$^{\prime\prime}$ \\
Synthesised beam size & 30$^{\prime\prime}$ \\
Noise & 2.8 K per 0.25 km s$^{-1}$ channel (10 hr) \newline
1.1 K per 1 km s$^{-1}$ channel (20 hr) \\
Threshold & 15 mJy \\
\hline
\end{tabular}
  \label{tab:proc-params}
\end{table}

\begin{figure*}
  \centering
  \includegraphics[width=\linewidth]{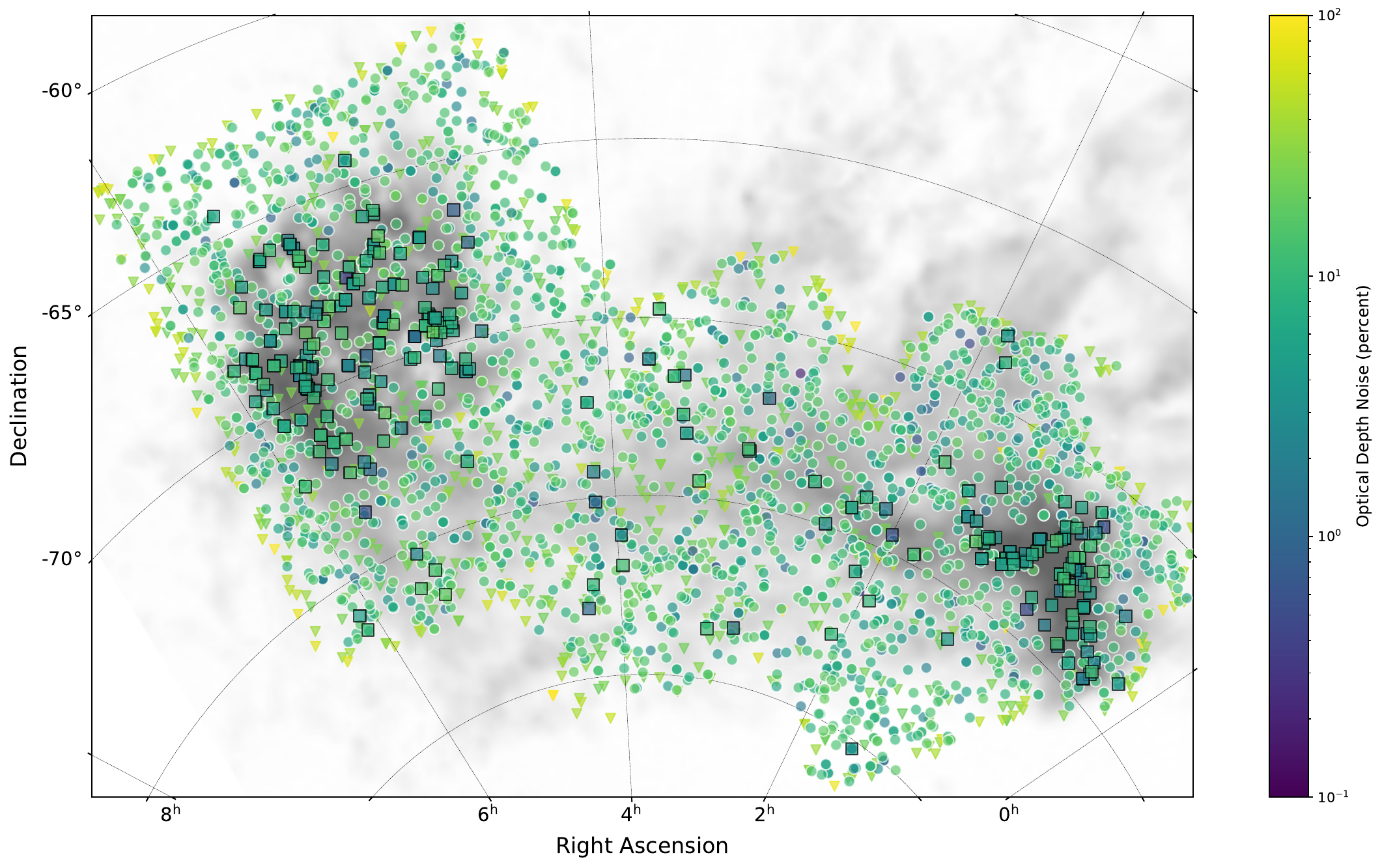}
  \caption{Map of all 3219 sightlines from GASKAP-HI pilot phase I and II observations of the Magellanic system where optical depth spectra were obtained. Each point is coloured by continuum noise level, with darker colours indicating lower noise. Squares are candidate detections (344), circles are non-detections (2116) and triangles are high noise spectra (759) excluded from further analysis. The background map is \hi\ column density from the GASS survey \citep{kalberla_haud_2015_gass}.}
  \label{fig:gaskap_pilot_abs_stacked}
\end{figure*}

In total, optical depth ($e^{-\tau}$) spectra were measured against 3219 unique continuum sources using the process described in \cite{dempsey_2022}.
In summary, in this process we image a small region around the source while excluding short baselines to produce a source cubelet with a typical $16^{\prime\prime} \times 14^{\prime\prime}$ synthesised beam, extract a 1D spectrum from this cubelet, and divide that spectrum by the continuum level as assessed in an offline velocity range to obtain the absorption spectrum in terms of $e^{-\tau}$.
The resulting coverage of continuum sources is shown in Figure \ref{fig:gaskap_pilot_abs_stacked} coloured by continuum noise level, $\sigma_\text{cont}$ (darker is lower noise).
Of these, 2460 spectra had $\sigma_\text{cont} \leq 0.2$ and were thus considered suitable for further analysis (shown as squares for candidate detections and circles for non-detections). 
The 759 excluded spectra (shown as triangles) are predominantly on the edges of the observed region where lower beam coverage has resulted in lower sensitivity. 
However, some of these are also located in regions of lower sensitivity where we only have Pilot Phase II observations, or there are gaps between the observations.

\begin{figure*}
  \centering
  \includegraphics[width=\linewidth]{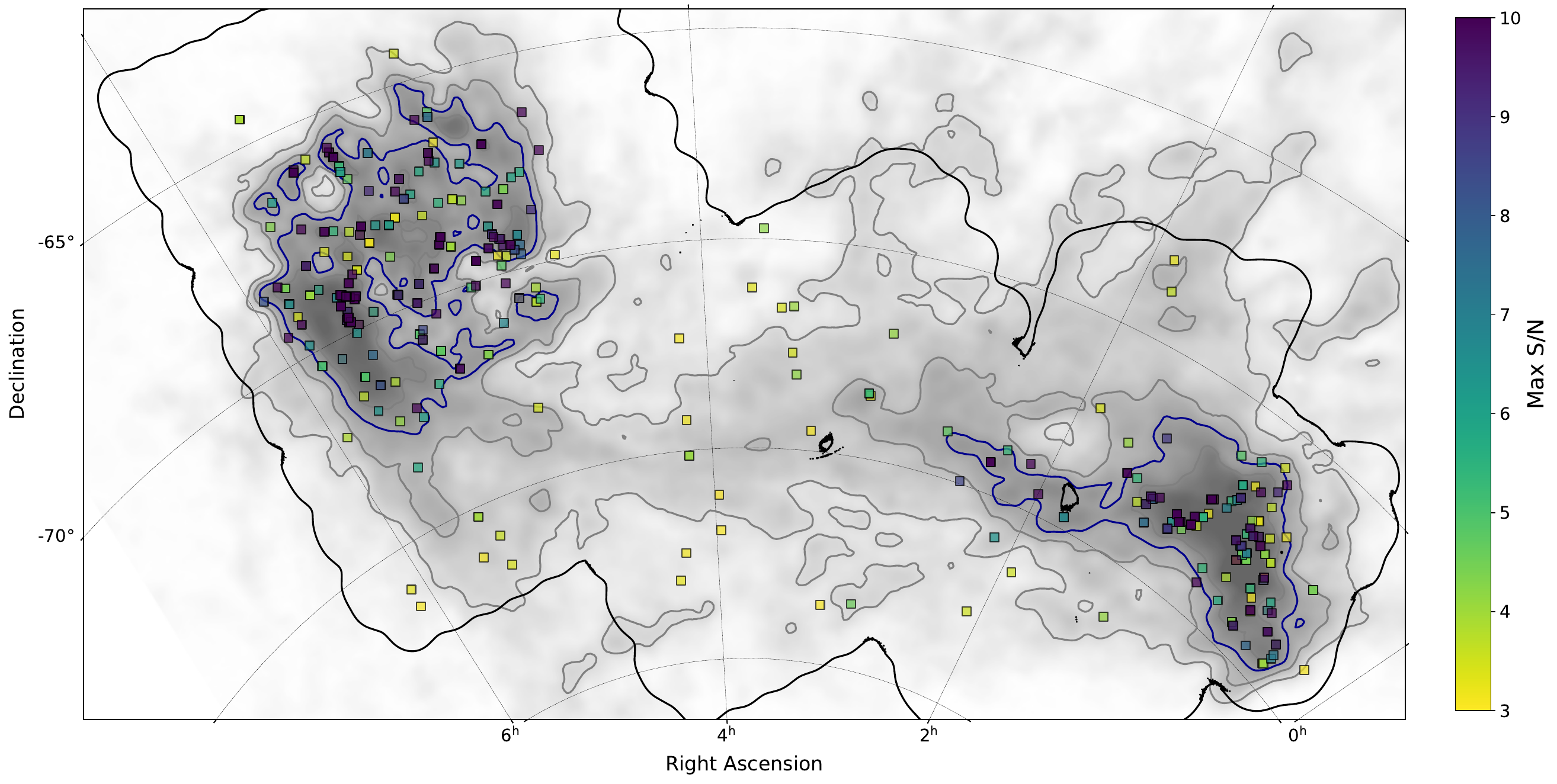}
  \caption{Spatial distribution of the 344 sightlines with candidate absorption detections at Magellanic velocities.
  Each candidate is coloured by its signal-to-noise ratio, with darker colours reflecting higher ratio.
  The background map is \hi\ column density from the GASS survey \citep{kalberla_haud_2015_gass}.
  Contours are shown at 0.18, 0.5, 1 and $2 \times 10^{21}$ cm$^{-2}$. The contour at $1 \times 10^{21} \textrm{ cm}^{-2}$ is shown in blue.
  The survey coverage is shown by the heavy black contour and also shows the two small coverage 'holes' noted in Section \ref{sec:data}.}
  \label{fig:gaskap_pilot_mag_detections}
\end{figure*}

\begin{table*}
\begin{threeparttable}
  \caption{Sample of the GASKAP-HI pilot II Magellanic spectrum catalogue. This is a sample of the key fields from the GASKAP-HI absorption spectrum catalogue for 13 sources. The full catalogue of all 2460 low-noise sources is available in the dataset \protect\cite{Dempsey2026}. }
\centering
\begin{tabular}{crrrrrrrr}
\hline \hline
Source & RA & Dec & Peak Flux\tnote{a} & $\sigma_{\rm cont}$\tnote{b} & Peak $\tau$\tnote{c} & N$_{\rm HI,uncorr}$\tnote{d} & $\langle T_{\rm S} \rangle$\tnote{e} & $f_\text{CNM,TS}$ \\
 & (deg) & (deg) & (mJy) &  &  & ($10^{21}$ cm$^{-2}$) & (K) &  \\
\hline
J014758$-$744831 & 26.9946 & $-74.809$ & 63 & 0.114 & $0.54\pm0.18$ & 2.0 & $305^{+160}_{-74}$ & $0.10^{+0.02}_{-0.05}$ \\
J014810$-$744345 & 27.0424 & $-74.729$ & 70 & 0.123 & $3.67\pm1.84$ & 2.0 & $102^{+16}_{-13}$ & $0.29^{+0.04}_{-0.05}$ \\
J015136$-$735424 & 27.9011 & $-73.907$ & 78 & 0.106 & $0.35\pm0.15$ & 0.8 & $201^{+160}_{-67}$ & $0.15^{+0.05}_{-0.12}$ \\
J015453$-$750829 & 28.7221 & $-75.142$ & 62 & 0.092 & $0.39\pm0.13$ & 1.0 & $479^{+1100}_{-1100}$ & $0.06^{+0.15}_{-0.14}$ \\
J015646$-$751412 & 29.1929 & $-75.237$ & 26 & 0.109 & $0.43\pm0.15$ & 0.8 & $101^{+29}_{-19}$ & $0.30^{+0.06}_{-0.08}$ \\
J015826$-$743823 & 29.6110 & $-74.640$ & 242 & 0.017 & $0.06\pm0.02$ & 0.9 & $2300^{+4000}_{-4700}$ & $0.01^{+0.03}_{-0.02}$ \\
J015952$-$743056 & 29.9694 & $-74.516$ & 845 & 0.006 & $0.14\pm0.01$ & 1.0 & $720^{+68}_{-62}$ & $0.04^{+0.00}_{-0.00}$ \\
J020635$-$740524 & 31.6472 & $-74.090$ & 64 & 0.051 & $0.32\pm0.08$ & 0.6 & $331^{+330}_{-120}$ & $0.09^{+0.03}_{-0.09}$ \\
J020638$-$735609 & 31.6619 & $-73.936$ & 200 & 0.016 & $0.25\pm0.02$ & 0.6 & $1560^{+2000}_{-760}$ & $0.02^{+0.01}_{-0.02}$ \\
J020640$-$743512 & 31.6698 & $-74.587$ & 26 & 0.055 & $0.29\pm0.08$ & 1.1 & $244^{+67}_{-42}$ & $0.12^{+0.02}_{-0.03}$ \\
J020645$-$743508 & 31.6877 & $-74.586$ & 24 & 0.072 & $0.31\pm0.11$ & 1.1 & $648^{+990}_{-380}$ & $0.05^{+0.03}_{-0.07}$ \\
\hline
\end{tabular}
 
\textbf{Notes:}
\begin{tablenotes}
\footnotesize
\item [a] Peak flux density of the source from the Selavy continuum catalogue.
\item [b] 1$\sigma$ continuum noise level of the absorption spectrum ($e^{-\tau}$). Does not include per channel emission noise.
\item [c] The maximum $\tau$ value within the Magellanic velocity range. Where the spectrum is saturated ($e^{-\tau} \le 0$) a minimum limit is specified based on the $1 \sigma$ noise level of the channel in the spectrum.
\item [d] Column density towards the continuum source, excluding Milky Way velocities, from the GASKAP-HI emission data under the assumption that the \hi\ is optically thin.
\item [e] The density-weighted mean spin temperature of the sight-line (see Section \ref{sec:compare_approach}).
\end{tablenotes}
  \label{tab:spectra-cat}
  \end{threeparttable}
\end{table*}

\begin{table*}
  \begin{threeparttable}
\caption{Sample of the GASKAP-HI pilot II Magellanic absorption feature catalogue. This is a sample of the key fields for those \hi\ absorption features detected in the spectra listed in Table \ref{tab:spectra-cat}. Note that multiple features are detected in some of the spectra, while other spectra have no detectable features. The full catalogue of all 1135 features is available in the dataset \protect\cite{Dempsey2026}.}
\centering
\begin{tabular}{llrrrrrrr}
\hline \hline
Source & Feature & Min Velocity & Max Velocity & Width\tnote{a} & Peak\tnote{b} & Peak $\tau$\tnote{c} & Significance\tnote{d} & Equiv. Width\tnote{e} \\
 &  & (kms$^{-1}$) & (kms$^{-1}$) & (kms$^{-1}$) & Absorption &  &  & (kms$^{-1}$) \\
\hline
J014810$-$744345 & J014810$-$744345\_143 & $143.9$ & $141.0$ & 4 & $0.88\pm0.15$ & $2.13\pm0.81$ & 6.0 & $4.57\pm0.31$ \\
J014810$-$744345 & J014810$-$744345\_169 & $169.4$ & $167.4$ & 3 & $0.97\pm0.13$ & $3.67\pm1.84$ & 7.2 & $6.25\pm0.25$ \\
J014810$-$744345 & J014810$-$744345\_183 & $183.0$ & $182.1$ & 2 & $0.61\pm0.14$ & $0.94\pm0.31$ & 4.3 & $1.52\pm0.21$ \\
J015826$-$743823 & J015826$-$743823\_9 & $9.1$ & $7.2$ & 3 & $0.11\pm0.02$ & $0.11\pm0.02$ & 5.5 & $0.31\pm0.03$ \\
J015952$-$743056 & J015952$-$743056\_11 & $11.0$ & $6.1$ & 6 & $0.10\pm0.01$ & $0.11\pm0.01$ & 15.1 & $0.34\pm0.02$ \\
J015952$-$743056 & J015952$-$743056\_151 & $151.7$ & $147.8$ & 5 & $0.13\pm0.01$ & $0.14\pm0.01$ & 18.9 & $0.44\pm0.02$ \\
J015952$-$743056 & J015952$-$743056\_2 & $2.2$ & $1.2$ & 2 & $0.04\pm0.01$ & $0.04\pm0.01$ & 5.7 & $0.07\pm0.01$ \\
J020638$-$735609 & J020638$-$735609\_141 & $141.8$ & $137.9$ & 5 & $0.22\pm0.02$ & $0.25\pm0.02$ & 13.3 & $0.76\pm0.04$ \\
J020640$-$743512 & J020640$-$743512\_9 & $10.0$ & $9.0$ & 2 & $0.22\pm0.06$ & $0.25\pm0.08$ & 3.6 & $0.45\pm0.09$ \\
\hline
\end{tabular}
 
\textbf{Notes:}
\begin{tablenotes}
\footnotesize
\item [a] The number of 1 km s$^{-1}$ channels the feature spans.
\item [b] The maximum measured absorption ($1-e^{-\tau}$) of the feature with the uncertainty in absorption spectrum at that velocity, values greater than 1 are saturated.
\item [c] The maximum $\tau$ value of the feature. Where the spectrum is saturated ($e^{-\tau} \le 0$) a minimum limit is specified based on the noise level in the peak channel.
\item [d] The highest single-channel significance of the feature as measured in absorption ($e^{-\tau}$).
\item [e] The integral of $\tau$ for the feature.  This will be a lower limit for saturated spectra as we use a minimum limit based on noise for each saturated velocity channel in the same way as peak $\tau$ above.
\end{tablenotes}
  \label{tab:abs-cat}
\end{threeparttable}
\end{table*}

For our dataset of Magellanic \hi\ absorption we have chosen to include only absorption features with v$_\text{LSRK} \ge 90$ km s$^{-1}$.
This baseline is driven by the SMC, which has emission located in the range 90 -- 190 km s$^{-1}$ \citep{pingel_2022}.
However, \cite{Richter_2015} found that clouds in the direction of the LMC at v$_\text{LSRK} = $ 150 km s$^{-1}$ were associated with the MW halo rather than the LMC.
So we caution that this dataset will still contain some absorption features that are physically located in the MW halo.
When working with detections at lower velocities and away from the main Magellanic bodies, we recommend an approach such as \cite{Poudel_2025} where each feature is examined to determine if they are red shifted MW gas or blue shifted Magellanic gas.

In many cases, sources may have been observed more than once across the pilot observations.
This may be the case either for repeated observations (e.g. phase I and II SMC observations), or for sources near the edges of fields and thus within the footprint of multiple observations.
In these cases we have stacked the cubelets for the source from each observation to produce a single cubelet with improved sensitivity.
When stacking, the lowest noise cubelet is chosen as the reference and the data for the other cubelets are interpolated onto the reference cubelet's velocity axis. 
The reference and interpolated cubelets are each assigned a weight using the inverse square of their continuum noise level:
\begin{equation}
w_i = \frac{1/\sigma_\text{cont,i}^2}{\sum^0_n(1/\sigma_\text{cont,n}^2)},
\label{eq:weighting}
\end{equation}
and the weighted data then summed.
An optical depth spectrum is then extracted from the combined cubelet and assessed as described in \cite{dempsey_2022}.
Where the continuum noise of the stacked spectrum is greater than that of the reference cubelet, then the stacked cubelet is discarded and the reference cubelet and its optical depth spectrum are used instead.
Further detail on this stacking process is provided in Appendix \ref{sec:stacking-app}.

Potential absorption was detected at Magellanic velocities (v$_\text{LSRK} \geq 90$ km s$^{-1}$) in 344 of our optical depth spectra.
A potential absorption detection is identified by a channel with $3\sigma$ or greater absorption and an adjacent channel with at least $2.8\sigma$ absorption.
More details on these criteria can be found in \cite{dempsey_2022}.
The locations of these potential detections, along with the significance of the peak absorption at Magellanic velocities (with darker colours showing higher significance), are shown in Figure \ref{fig:gaskap_pilot_mag_detections}.
Of these candidate detections, 79 (or 23\%) have absorption significance $< 4\sigma$.
For these spectra with a lower absorption significance, there is a risk that the detection may not be sufficiently significant to confirm in the current data.
To mitigate this risk, these lower significance candidate detections should be checked before use for adjacent noise in the optical depth spectrum and against emission in the same location to identify whether they are likely to be confirmed detections.

A subset of sources are described in Table \ref{tab:spectra-cat}, with the full set of sources provided in the dataset. 
We also provide a subset of absorption features in Table \ref{tab:abs-cat}, with the full set of features provided in the dataset. 

\begin{table}
\caption{Source detections statistics for each major component of the Magellanic system.
The SMC is bounded by a simplified contour where $N_{\mathrm{H\,I}} \geq 2 \times 10^{21}$ cm$^{-2}$. 
The LMC is bounded by a simplified contour where $N_{\mathrm{H\,I}} \geq 5 \times 10^{20}$ cm$^{-2}$. 
The Bridge is defined as the region within 1$^h$30$^m$ $< \alpha <$ 4$^h$30$^m$ and $-80^{\circ}$ $< \delta < -70^{\circ}$.
Other includes any sources located outside these areas in the survey.}
\centering
\begin{tabular}{ l r r r r r}
\hline \hline
 Component & Area & Num & Num & Det. & Mean \\ 
  &  & Sources & Det. & Rate & $\sigma_\text{cont}$ \\ 
  & (deg$^2$) &  &  & (deg$^{-2}$)& \\ 
 \hline
 SMC & $10.5$ & $117$ & $85$ &  $8.1$ &  $0.069$  \\ 
 Bridge & $119.8$ & $793$ & $35$ &  $0.3$ &  $0.092$  \\ 
 LMC & $57.0$ & $438$ & $192$ &  $3.4$ &  $0.091$  \\ 
 Other & -- & $1112$ & $32$ &  -- & $0.096$  \\ 
\hline
\end{tabular}
 
  \label{tab:mag_det_stats}
\end{table}

\begin{figure}
  \centering
  \includegraphics[width=\linewidth]{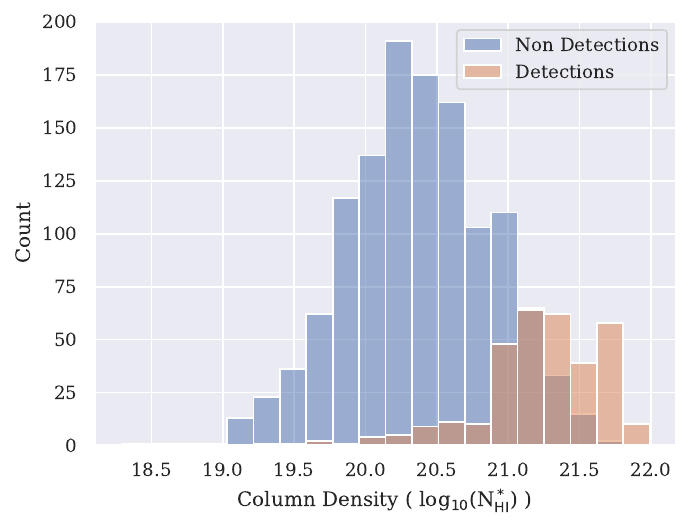}
  \caption{Distribution by \hi\ column density of sightlines both with and without candidate absorption detections.
  The GASKAP-HI column density shown here is measured under the assumption that the \hi\ is optically thin. The blue bars show the count of spectra without detected absorption features in each column density bin, while the orange bars show the count of spectra with absorption features in those bins.}
  \label{fig:detections_nh_hist}
\end{figure}

With coverage of the entirety of both Magellanic Clouds as well as the area between them, we can compare the chances of detecting absorption within the Clouds to the lower density regions surrounding and between them.
We find that most detections are in the Clouds rather than the Bridge and other areas outside the Clouds. 
Table \ref{tab:mag_det_stats} shows that the detection rate is highest in the compact dense SMC, still common in the larger, less compact LMC and relatively rare in the sparse Bridge.
Correspondingly, we find that the candidate absorption detections are much more common in sightlines of higher \hi\ column density, as shown in Figure \ref{fig:detections_nh_hist}.
Detections are only more common than non detections once $N_{\mathrm{H\,I}} > 10^{21}$ cm$^{-2}$.

Notably, in Figure \ref{fig:detections_nh_hist} we can see a population of absorption detections in low \hi\ column density regions though. 
The majority of these are on the outskirts of the two Clouds.
A sample of these are examined in detail in \cite{chen_2025}.
There are also some detections in low density regions of the Bridge and these are discussed in Section \ref{sec:bridge-abs}.

For the subset of spectra towards extra-galactic sources with detections in the Magellanic Bridge region (defined in section \ref{sec:bridge-abs}) we have fitted the spectra with Gaussians representing individual gas clouds along the line of sight.
We jointly decompose the absorption and emission spectra using the technique described in \cite{nguyen_2024}.
In summary, we take 20 samples from the GASKAP-HI emission data of the emission near the source, but not within a beam of the source.
Each sample is separated by 30$^{\prime\prime}$ and any samples that show signs of absorption contamination are excluded.
Following the \citetalias{heiles_millennium_2003-1} fitting methodology, we first fit the absorption spectrum and then fit each emission spectrum with Gaussians for each absorption component and then multiple emission components.
However, in our case, given the small number of spectra to be fitted, we chose to provide initial values for the Gaussians to fit the emission spectrum for each target.
From these fits we calculated the cold gas fraction ($f_\textnormal{CNM} = \sum N_\textnormal {\hi,CNM} / (\sum N_\textnormal {\hi,CNM} + \sum N_\textnormal {\hi,WNM}) $).
An example fit to the emission and absorption spectra is shown in Figure \ref{fig:J013218-715348} and plots for all fits are provided in Appendix \ref{sec:bridge-spectra-plots}.
These spectra are discussed in more detail in section \ref{sec:bridge-abs}.

\begin{figure}
  \centering
  \includegraphics[width=\linewidth]{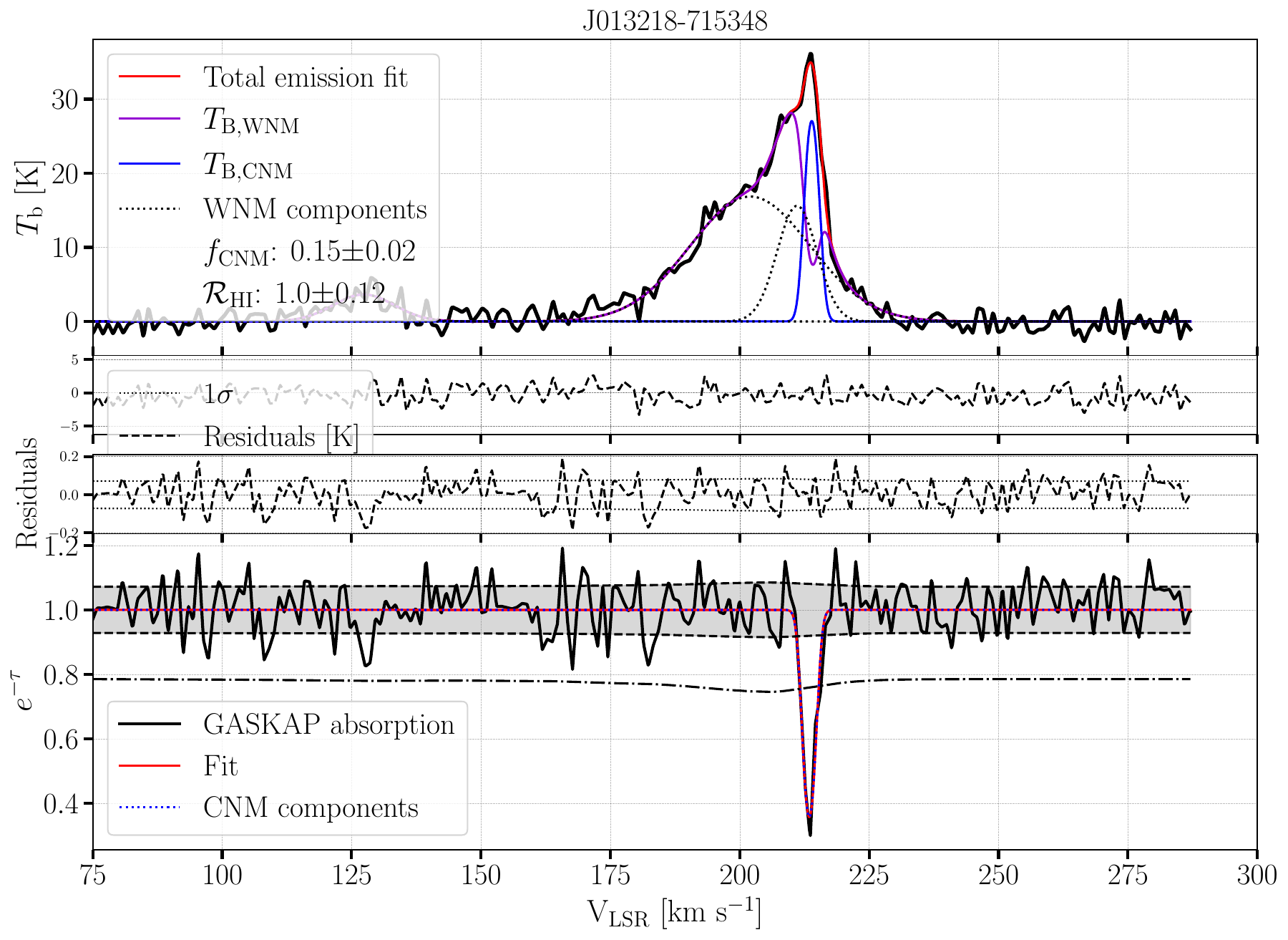}
  \caption{
  Gaussian decomposition of \hi\ emission and absorption spectra for J013218$-$715348.
  The lower section shows the absorption spectrum ($e^{-\tau}$) in black with the fit shown in red.
  The $1\sigma$ noise envelope is shown as gray shading and the $3 \sigma$ threshold is shown as a dot-dashed line.
  A residual between the fit and the spectrum is shown at the top of the lower panel.
  The upper panel shows the brightness temperature ($T_b$) spectrum in black with the fit shown as the red line.
  The individual WNM components are shown as dotted lines and the CNM component is shown as a blue line.
  Again the residual between the fit and the emission spectrum is shown at the bottom of the top panel. The full set of Bridge spectra are shown in Appendix \ref{sec:bridge-spectra-plots}.}
  \label{fig:J013218-715348}
\end{figure}

\section{Cold \hi\ gas in the Magellanic Bridge}
\label{sec:bridge-abs}

\begin{figure*}
  \centering
  \includegraphics[width=\linewidth]{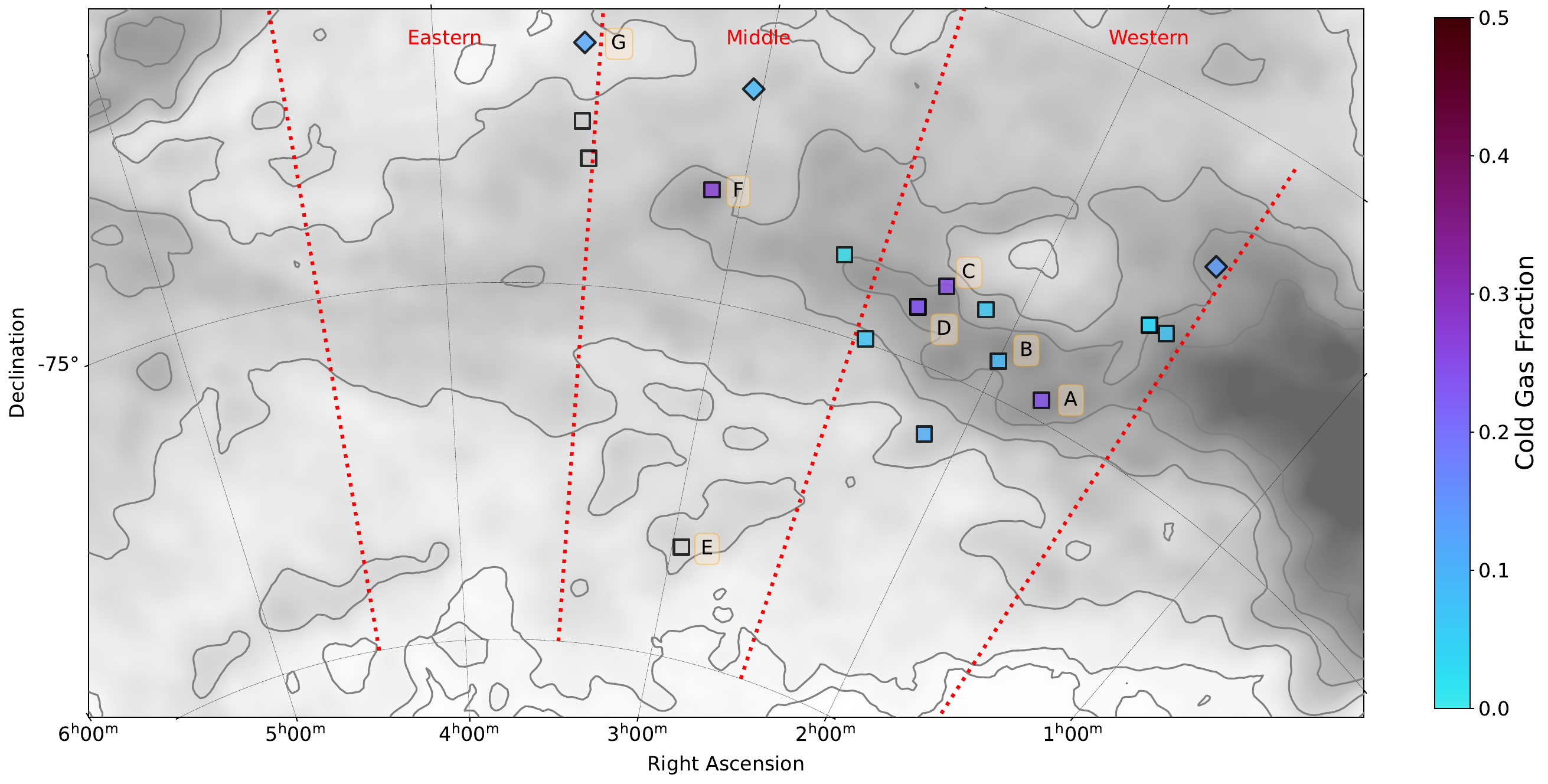}
  \caption{Map of cold gas detections in absorption in the MB region. The 11 filled squares (2 of which represent a source pair each) and 3 filled diamonds are detections at Magellanic velocities, coloured by the cold gas fraction of the sight line. 
  The cold gas fraction is determined from Gaussian fitting of each spectrum. 
  The diamonds indicate sightlines where the cold gas components have $v > 200$ km s$^{-1}$.
  The 3 empty squares show the positions of detections that could not be Gaussian decomposed.
  Seven sources discussed in the text are marked with letters (A--G).
  Red dotted lines delineate the western, middle and eastern regions of the Bridge.
  Note: 19 candidate detections with $S/N < 4$ are not shown but can be seen in Figure \ref{fig:gaskap_pilot_mag_detections}.
  The background map is \hi\ column density from the GASS survey \citep{kalberla_haud_2015_gass}.}
  \label{fig:bridge_cold_gas}
\end{figure*}

We can use the \hi\ absorption observations of the Magellanic Bridge as a case study for the overall survey of the Magellanic system.
The Magellanic Bridge is a field of debris between the SMC and the LMC, stretching from the "Wing" of the SMC \citep{shapley_1940} to the arm B of the LMC \citep{staveley-smith_2003}, with a \hi\ mass of around half that of the SMC \citep{bruns_2005}.
Here we analyse the Right Ascension range 1$^h$30$^m$ $< \alpha <$ 4$^h$30$^m$ of the Bridge.
In the GASKAP-HI observations we have detected 19 sources with $4\sigma$ or better absorption in the MB region.
Of these we have been able to Gaussian decompose 16 spectra, which are shown as squares coloured by cold gas fraction in Figure \ref{fig:bridge_cold_gas} and are listed in Table \ref{tab:bridge-gaussian}.
These include two source pairs J013701$-$730415 / J013704$-$730412 and J021936$-$741604 / J021939$-$741558.
Three sources, J025721$-$782714, J033046$-$731238, and J033228$-$724120, were not decomposed as the emission signal-to-noise ratio was too low ($S/N_{T_b} < 7$).
These are shown as empty squares in the figure.
As with the overall system, there is a strong correlation between the denser \hi\ gas density and the detection of absorption.

\begin{table*}
  \caption{Catalogue of cold (Cx) and warm (Wx) gas Gaussian components of each spectrum with a \hi\ absorption detection in the Bridge.}
\centering
\begin{tabular}{ c c c c c r r r r r }
 Source & RA & Dec & S/N & Id & $\tau_\mathrm{peak}$ & $T_\mathrm{b,peak}$ & Central Velocity & FWHM & $T_\mathrm{spin}$ \\ 
  &  (deg) & (deg) & & & & (K) & (km.s$^{-1}$) & (km.s$^{-1}$) & (K) \\ 
 \hline
 J013218$-$715348 & 23.0784 & $-$71.8967 & 38 & C0 & $1.03 \pm 0.03$ & & $213.42 \pm 0.04$ & $2.61 \pm 0.09$ & $28.2 \pm 9.3$  \\ 
 & & & & W0 & & $16.90 \pm 2.44$ & $201.70 \pm 2.29$ & $26.22 \pm 2.82$ & \\ 
 & & & & W1 & & $15.98 \pm 4.20$ & $211.20 \pm 0.80$ & $10.54 \pm 2.46$ & \\ 
 & & & & W2 & & $3.49 \pm 1.07$ & $128.74 \pm 2.92$ & $20.39 \pm 7.38$ & \\ 
 \hline
 J013329$-$730304 & 23.3734 & $-$73.0513 & 38 & C0 & $0.25 \pm 0.01$ & & $157.51 \pm 0.11$ & $3.93 \pm 0.26$ & $49.1 \pm 10.5$  \\ 
 & & & & W0 & & $15.29 \pm 0.81$ & $163.79 \pm 0.96$ & $33.61 \pm 2.51$ & \\ 
 & & & & W1 & & $10.82 \pm 1.00$ & $195.87 \pm 0.81$ & $15.79 \pm 1.91$ & \\ 
 \hline
 J013701$-$730415 & 24.2569 & $-$73.0711 & 13 & C0 & $0.27 \pm 0.01$ & & $190.23 \pm 0.04$ & $2.63 \pm 0.10$ & $16.3 \pm 6.0$  \\ 
 & & & & W0 & & $8.07 \pm 0.54$ & $166.62 \pm 2.39$ & $32.81 \pm 4.25$ & \\ 
 & & & & W1 & & $6.38 \pm 1.02$ & $193.86 \pm 2.03$ & $22.07 \pm 3.67$ & \\ 
 \hline
 J013704$-$730412 & 24.2694 & $-$73.0702 & 15 & C0 & $0.36 \pm 0.01$ & & $189.90 \pm 0.05$ & $3.45 \pm 0.12$ & $12.4 \pm 5.0$  \\ 
 & & & & W0 & & $7.49 \pm 0.87$ & $177.62 \pm 2.29$ & $29.38 \pm 3.96$ & \\ 
 & & & & W1 & & $6.87 \pm 0.91$ & $178.60 \pm 2.28$ & $25.00 \pm 4.22$ & \\ 
 \hline
 J014810$-$744345 & 27.0424 & $-$74.7294 & 22 & C0 & $1.57 \pm 0.06$ & & $142.53 \pm 0.05$ & $2.74 \pm 0.11$ & $27.3 \pm 14.8$  \\ 
 & & & & C1 & $2.92 \pm 0.06$ & & $168.05 \pm 0.02$ & $1.68 \pm 0.04$ & $10.3 \pm 5.3$  \\ 
 & & & & C2 & $0.85 \pm 0.06$ & & $182.83 \pm 0.07$ & $2.02 \pm 0.17$ & $25.3 \pm 8.4$  \\ 
 & & & & W0 & & $31.24 \pm 2.89$ & $140.55 \pm 0.44$ & $14.03 \pm 1.18$ & \\ 
 & & & & W1 & & $20.12 \pm 1.63$ & $176.56 \pm 0.89$ & $23.44 \pm 2.21$ & \\ 
 \hline
 J015952$-$743056 & 29.9694 & $-$74.5158 & 17 & C0 & $0.14 \pm 0.00$ & & $149.38 \pm 0.03$ & $3.06 \pm 0.06$ & $95.5 \pm 14.4$  \\ 
 & & & & W0 & & $9.76 \pm 0.63$ & $147.41 \pm 1.05$ & $31.52 \pm 2.73$ & \\ 
 & & & & W1 & & $7.99 \pm 0.64$ & $184.43 \pm 1.04$ & $21.65 \pm 2.48$ & \\ 
 \hline
 J020638$-$735609 & 31.6619 & $-$73.9361 & 7 & C0 & $0.25 \pm 0.01$ & & $139.57 \pm 0.04$ & $2.99 \pm 0.08$ & $14.2 \pm 6.4$  \\ 
 & & & & W0 & & $9.18 \pm 0.43$ & $151.85 \pm 0.51$ & $20.39 \pm 1.27$ & \\ 
 \hline
 J020752$-$755242 & 31.9692 & $-$75.8784 & 10 & C0 & $0.54 \pm 0.03$ & & $141.79 \pm 0.07$ & $2.38 \pm 0.16$ & $12.7 \pm 3.2$  \\ 
 & & & & W0 & & $5.16 \pm 0.45$ & $141.03 \pm 0.70$ & $17.37 \pm 1.75$ & \\ 
 & & & & W1 & & $2.10 \pm 0.23$ & $174.49 \pm 2.70$ & $36.71 \pm 7.14$ & \\ 
 \hline
 J021541$-$735103 & 33.9239 & $-$73.8509 & 14 & C0 & $0.48 \pm 0.03$ & & $172.37 \pm 0.13$ & $4.82 \pm 0.30$ & $61.8 \pm 7.2$  \\ 
 & & & & W0 & & $10.49 \pm 1.51$ & $158.86 \pm 0.78$ & $14.67 \pm 2.02$ & \\ 
 & & & & W1 & & $9.63 \pm 1.08$ & $169.27 \pm 1.64$ & $34.28 \pm 3.75$ & \\ 
 \hline
 J021936$-$741604 & 34.9020 & $-$74.268 & 23 & C0 & $1.25 \pm 0.04$ & & $157.55 \pm 0.03$ & $2.26 \pm 0.08$ & $21.4 \pm 9.9$  \\ 
 & & & & W0 & & $21.38 \pm 1.50$ & $161.68 \pm 0.40$ & $12.64 \pm 1.00$ & \\ 
 & & & & W1 & & $18.54 \pm 1.40$ & $169.70 \pm 0.46$ & $14.30 \pm 1.21$ & \\ 
 \hline
 J021939$-$741558 & 34.9155 & $-$74.2664 & 28 & C0 & $1.38 \pm 0.01$ & & $157.44 \pm 0.01$ & $2.63 \pm 0.02$ & $20.7 \pm 8.5$  \\ 
 & & & & C1 & $0.49 \pm 0.01$ & & $175.48 \pm 0.03$ & $2.84 \pm 0.07$ & $24.5 \pm 9.0$  \\ 
 & & & & W0 & & $19.71 \pm 1.77$ & $153.91 \pm 0.48$ & $13.28 \pm 1.23$ & \\ 
 & & & & W1 & & $13.43 \pm 1.43$ & $177.62 \pm 0.76$ & $17.81 \pm 2.12$ & \\ 
 \hline
 J022741$-$745621 & 36.9210 & $-$74.9394 & 11 & C0 & $0.26 \pm 0.01$ & & $157.25 \pm 0.05$ & $1.96 \pm 0.12$ & $16.4 \pm 5.7$  \\ 
 & & & & W0 & & $5.66 \pm 0.34$ & $159.97 \pm 0.57$ & $21.96 \pm 1.45$ & \\ 
 \hline
 J023712$-$735417 & 39.3009 & $-$73.9049 & 23 & C0 & $0.50 \pm 0.04$ & & $199.33 \pm 0.07$ & $1.56 \pm 0.16$ & $8.3 \pm 5.3$  \\ 
 & & & & W0 & & $8.33 \pm 0.82$ & $174.65 \pm 1.40$ & $26.30 \pm 2.94$ & \\ 
 & & & & W1 & & $6.31 \pm 2.29$ & $167.90 \pm 1.88$ & $13.96 \pm 3.48$ & \\ 
 & & & & W2 & & $4.87 \pm 2.12$ & $167.90 \pm 1.83$ & $6.65 \pm 3.70$ & \\ 
 & & & & W3 & & $3.37 \pm 0.61$ & $228.01 \pm 1.86$ & $19.93 \pm 4.72$ & \\ 
 \hline
 J030138$-$715634 & 45.4102 & $-$71.9428 & 9 & C0 & $0.04 \pm 0.00$ & & $212.01 \pm 0.16$ & $4.74 \pm 0.38$ & $47.0 \pm 24.0$  \\ 
 & & & & W0 & & $5.00 \pm 0.52$ & $211.59 \pm 0.59$ & $19.53 \pm 1.73$ & \\ 
 \hline
 J030600$-$732644 & 46.5023 & $-$73.4456 & 25 & C0 & $1.18 \pm 0.06$ & & $156.96 \pm 0.07$ & $2.73 \pm 0.17$ & $25.8 \pm 8.6$  \\ 
 & & & & C1 & $1.10 \pm 0.06$ & & $161.37 \pm 0.07$ & $2.70 \pm 0.18$ & $14.1 \pm 7.7$  \\ 
 & & & & W0 & & $14.81 \pm 3.51$ & $159.96 \pm 1.45$ & $13.25 \pm 2.13$ & \\ 
 & & & & W1 & & $13.32 \pm 3.96$ & $152.77 \pm 0.34$ & $4.09 \pm 1.25$ & \\ 
 & & & & W2 & & $3.26 \pm 0.64$ & $222.88 \pm 4.06$ & $37.41 \pm 11.47$ & \\ 
 & & & & W3 & & $5.55 \pm 0.92$ & $187.98 \pm 1.50$ & $16.69 \pm 3.72$ & \\ 
 \hline
 J033252$-$713453 & 53.2177 & $-$71.5816 & 8 & C0 & $0.07 \pm 0.01$ & & $253.25 \pm 0.14$ & $3.87 \pm 0.34$ & $30.0 \pm 9.4$  \\ 
 & & & & W0 & & $2.23 \pm 0.24$ & $257.57 \pm 1.18$ & $29.01 \pm 3.16$ & \\ 
 \hline
\end{tabular}
 
  \label{tab:bridge-gaussian}
\end{table*}

\begin{table}
  \caption{Statistics for each of the Magellanic Bridge regions used in analysis in Section \ref{sec:bridge-abs}.
  The Num Fitted column shows the number of sight-lines with absorption detections that were successfully Gaussian decomposed.
  The $\sigma_\text{cont}$ column shows the mean continuum noise level across all sources in the region.
  The $\tau$ limit is the lowest $\tau$ value that could be detected at 3$\sigma$ significance across all sources in the region.}
\centering
\begin{tabular}{ l r r r r r}
\hline \hline
  &  & Absorption & Num & Mean & Mean \\ 
 Region & Sources & Detections & Fitted & $\sigma_\text{cont}$ & $\tau$ Limit \\ 
 \hline
 Western & $275$ & $12$ & $12$ & $0.089$ &  $0.33$  \\ 
 Middle & $225$ & $4$ & $3$ & $0.090$ &  $0.34$  \\ 
 Eastern & $221$ & $3$ & $1$ & $0.099$ &  $0.39$  \\ 
\hline
\end{tabular}
  \label{tab:mb_regions}
\end{table}

The Gaussian decomposition derived spin temperatures across the Bridge are generally very low, ranging from 8 K to 49 K, with two outliers at 62 K (at J021541$-$735103, symbol C in Figure \ref{fig:bridge_cold_gas}) and 96 K (at J015952$-$743056, B in Figure \ref{fig:bridge_cold_gas}) .
The median spin temperature is 23 K.  These values are lower than the typical spin temperature in the SMC of $\sim 30$ K, as found by \cite{jameson_2019} and significantly lower than typical values in the Milky Way  (50 -- 100 K) \citep{mcclur23}.
The values are more consistent with the spin temperatures found in the outskirts of the SMC and LMC by \cite{chen_2025}.
These lower gas temperatures are fully consistent with the expectations for low metallicity environment of the Magellanic Bridge (\citealt{lehner_2008} find a metallicity of [Z/H] $= -1.05 \pm 0.06$ dex solar, about a factor of two less than the SMC, however this value likely varies across the Bridge as \citealt{misawa_2009} find a [Z/H] of $-1.0$ to $-0.5$ dex solar towards PKS 0312$-$770).
Models of heating and cooling at low metallicity by \cite{bialy_2019} show that the typical \hi\ temperature is lower than at solar metallicity.
We speculate that these spin temperatures are even lower than  in the SMC because the metallicity in the Bridge environment is even lower metallicity than in the main body of the SMC.  

We can divide the MB into three arbitrary divisions of equal size, near the SMC (western), middle of the Bridge, and far from the SMC (eastern), as shown by the red dotted lines in Figure \ref{fig:bridge_cold_gas}. 
In the region closest to the SMC (1$^h$30$^m$ $< \alpha <$ 2$^h$30$^m$) we have 12 of the 16 decomposed spectra with detections.
While most of the spectra have a single absorption instance, close to the SMC we find J014810$-$744345 (A in Figure \ref{fig:bridge_cold_gas}) and J021939$-$741558 (D in Figure \ref{fig:bridge_cold_gas}) which have three and two absorption components respectively.
Within this region we also see a strong association of cold gas detections with the higher \hi\ column density regions of the Bridge, although we also have two detections in diffuse gas to the south of the main \hi\ tail.
\cite{kim_2025} (submitted) explored cold \hi\ gas in detail around two shells in this region using GASKAP-HI emission data, finding that the cold gas exists in a wide range of scales in the Bridge.
Interestingly, there is no general trend of cold gas fraction with distance along the Bridge and the fraction seems more influenced by local conditions.
Based on the Gaussian fitted data, we find a $\overline{f_\text{CNM}} = 0.12 \pm 0.08$.
This is very similar to the value for the SMC found in \cite{dempsey_2022}, noting that the SMC value is based on mean spin temperatures.

This western region also contains the two sightlines where \cite{kobulnicky_1999} reported peak optical depths of $\tau = 0.03$ and $\tau = 0.10$ in B0202$-$765 and B0312$-$770 respectively. 
We did not detect absorption towards these sightlines and have established an upper limit to optical depth in B0202$-$765 and B0312$-$770 of $\tau < 0.018$ and $\tau < 0.031$ respectively based on the $4\sigma$ limits from the noise in the GASKAP-HI spectra.
Working in conjunction with the original authors to re-examine the Australia Telescope Compact Array data, we found that the absorption is only present on the shorter baseline, which indicates that the result may have been confusion by large-scale emission structure.

The middle of the Bridge region (2$^h$30$^m$ $< \alpha <$ 3$^h$30$^m$) contains three of the remaining detections of cold gas. 
With a lower \hi\ column density level than the western region, we expect there will be less cold \hi\ in this middle region than the western region.
Surprisingly this region also includes the highest cold gas fraction of our sample, with J030600$-$732644 (F in Figure \ref{fig:bridge_cold_gas}) having a cold gas fraction of $\approx26\%$. 
This particular source is on the edge of a large shell surrounding a complex of young stars \citep{Mackey_2017} and is discussed further in the next section.

In the eastern region (3$^h$30$^m$ $< \alpha <$ 4$^h$30$^m$), farthest from the SMC and closest to the LMC, we find only three sightlines with significant \hi\ absorption, and are able to decompose only a single spectrum J033252$-$713453 (G in Figure \ref{fig:bridge_cold_gas}). 
This spectrum has the least emission (with a peak brightness temperature $T_{\rm b,peak} \sim 2.23$ K) of any spectrum we were able to decompose and is in a surprisingly diffuse environment (with an \hi\ column density $N_{\rm HI,uncorr} \sim 4.1 \times 10^{19}$ cm$^{-2}$).
There were also a handful of detections with marginal significance, hinting at the potential for further detections should more sensitive observations be taken in the future.
\cite{morelli_2025} recently reported the discovery of cold \hi\ absorption in the sight-line towards J033242.97$-$724904.5.
We see evidence of the same absorption in the GASKAP-HI data; however, the signal to noise level of our observation is not sufficient to be a detection in our data set.
When we compare the detection limits of cold \hi\ for the low-noise sightlines (see Table \ref{tab:mb_regions}) we find that the eastern region has 10\% higher noise (mean $\sigma_\text{cont}$) than the western region.
We also see that there are 20\% less low-noise sight-lines in the eastern region than the western.
These are both the result of the shorter dwell time per field for the pilot phase II observations than for the pilot phase I observations, as well as the gaps between phase I and phase II fields.
By scaling the noise of the spectra with detections in the western field we find that only a single detection would have been lost at the mean noise levels of the eastern field.
As a result, we can still be confident in stating that the optical depth of cold \hi\ is far lower in the eastern region than closer to the SMC.
We look forward to future deep GASKAP-HI observations of this region to further explore cold \hi\ at the LMC side of the Bridge.

Overall, we see a fall-off in the number of cold gas detections with distance from the SMC.
This supports the conclusion that the majority of cold gas has been either been pulled from the SMC in earlier interactions, along with the warm \hi\ seen in emission, or that the cold gas was formed as a part of those interactions that pulled the warm gas from the SMC.
However, we do not see a similar gradient in the cold gas fraction of the spectra with detections. 
We will discuss the variability of cold gas fraction further in the next section.

\subsection{A First Look at Cold \hi\ gas and the Structure of the Bridge}

With a spatial sample of cold \hi\ gas we can start to examine how the gas relates to the structure of the Bridge.
\cite{muller_2004} found that the Bridge was best described as two filamentary \hi\ structures, one stretching to the north of the low density region centred on ($\alpha = 2^h$ , $\delta = -73$\degr) and the other to the south of that region.
These filaments represent two arms of gas emanating from the SMC.
They identified the filaments by their distinct velocity ranges to the north and south of $\delta \approx 73$\degr $15^\prime$.
In our sample of cold gas detections, we find a similar split in velocity distribution of the cold gas.
The three northern sightlines (marked as diamonds in Figure \ref{fig:bridge_cold_gas}) are the only ones with cold gas components with $v > 200$ km s$^{-1}$.
In contrast, all of our cold gas detections south of $\delta \approx -73$\degr $15^\prime$ have a lower velocity ( $v < 200$ km s$^{-1}$).

With cold \hi\ gas being critical to the star formation process \citep[e.g.][]{krumholz_2012,glover_2012}, it can be instructive to compare the positions of the cold gas with those of stars formed within the Bridge.
The existence of a young stellar population within the Bridge is well known \citep[e.g.][]{irwin_1985}.
\cite{Belokurov_2017} traced two stellar bridges between the SMC and LMC, one of Young Main Sequence (YMS) stars and another of older RR Lyrae stars. 
We see cold \hi\ associated with the young stellar bridge, but little cold \hi\ coincident with the old stellar bridge.
The YMS bridge follows the outcrop of dense \hi\ from the SMC, while the older RR Lyrae bridge traces the leading edge of the \hi\, ($\delta > -79$\degr) well in advance of the dense \hi\ bridge (see their Fig 15).
\cite{Mackey_2017} found that many of the young star clusters were $\sim$30 Myr in age, consistent with them being formed in-situ in the Bridge.
We have good coverage of both of these regions, with the exception of a small area where the RR Lyrae bridge turns back  to the LMC ($\alpha \approx 5^h$, $\delta \approx -78$\degr).
We have detected cold gas in only one sight-line in the RR Lyrae bridge, J025721$-$782714 (E in Figure \ref{fig:bridge_cold_gas}).
In contrast, the majority of the cold gas we have mapped follows the YMS bridge, providing a reservoir for this star formation.

Given the larger number of cold gas detections noted earlier in the western region, closest to the SMC, we can surmise that much of this cool gas likely originates from the formation of the bridge, potentially when the cold gas was pulled from the SMC.
Equally however, the cool gas may have been formed during this process. 
The galaxy interaction that formed the bridge would trigger turbulence in the \hi\ \citep{bournaud_2010,duc_2013} and these turbulent flows can trigger the condensation of CNM \citep{audit_hennebelle_2005}.

Intriguingly, \cite{Mackey_2017} found triggered star formation in the edge of a \hi\ shell, with the shell stars 10$-$15 Myr younger than the stars in the core.
The high cold gas fraction we detect on the edge of this shell (against source J030600$-$732644, F in Figure \ref{fig:bridge_cold_gas}) likely points to enhancement or formation of cold gas in-situ as stellar feedback from the massive stars in the core of the shell compresses the gas on its outskirts.
The velocity structure of this cold gas is also intriguing, as it exhibits a double peak structure in absorption, unlike any other in the Bridge sample.
Similar in-situ enhancement of the cold gas due to local turbulence and star forming activity may explain the high variability of cold gas fraction across the Bridge.

\subsection{Comparison of mean spin temperature and Gaussian approaches}
\label{sec:compare_approach}

\begin{figure}
  \centering
  \includegraphics[width=\linewidth]{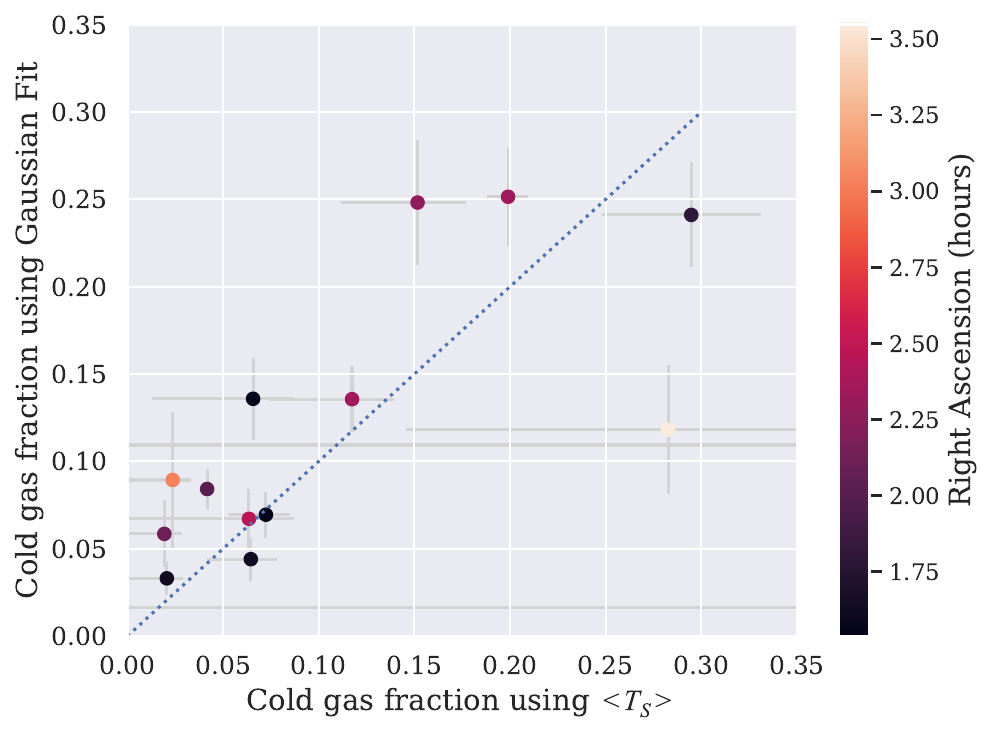}
  \caption{Comparison of cold gas fractions for each sightline as calculated using the mean spin temperature and Gaussian techniques. The diagonal dotted line shows equality. Right Ascension is used here as a proxy for distance from the SMC.}
  \label{fig:cold_gas_fraction_comparison}
\end{figure}

We have used a Monte-Carlo bootstrapping method to calculate density-weighted mean spin temperatures ($\langle T_{\rm S} \rangle$) and cold gas fractions ($f_\text{CNM,TS}$) across the Magellanic System \cite[see][section 5 for details]{dempsey_2022}.
Briefly, the density-weighted mean spin temperature is measured by comparing the integrals of brightness temperature and optical depth using the following formula \citep[Eq. 4]{dickey_2000}:
\begin{equation}
    \langle T_{\rm S} \rangle = \frac{\int T_{\rm B}(\text{v})~d\text{v}}{\int 1-e^{-\tau(\text{v})}~d\text{v} } = \int \frac{n(\text{s})}{[n(\text{s})/T_{\rm S}(\text{s})]} d\text{s} ,
\end{equation}
where $n(\text{s})$ is the line-of-sight volume density.
We calculate the cold gas fraction as $f_\text{CNM,TS} \simeq T_{\rm c}/\langle T_{\rm s} \rangle$ \citep[Eq. 7]{dickey_2000} where, following \citep{dempsey_2022} we assume $T_{\rm c} = 30$ K.
As discussed in Section \ref{sec:data}, within the Bridge we have also measured these quantities by fitting Gaussians to the emission and absorption spectra.
Comparing the outcomes of these two techniques (see Figure \ref{fig:cold_gas_fraction_comparison}) we note that four are consistent within uncertainties, three have a lower value from the Gaussian approach, and six have a higher value in the Gaussian derived value.
Overall, we find that they produce generally compatible results for the spectra in our MB sample.

Both approaches struggle to analyse noisy emission spectra in areas of faint emission.
An example is J033252$-$713453  where the \FCNM\ calculated with the two methods differ significantly ($f_\text{CNM,Gaussian}$ $= 0.12 \pm 0.04$ while $f_\text{CNM,TS}$ $= 0.29 \pm 0.15$).
While the absorption component detected here is a simple single Gaussian, the region in the direction of this continuum source has quite faint and variable emission ($T_\text{b} \le 4$ K).
In addition, the emission spectrum is at the lower edge of our signal to noise criteria. 
For the Gaussian fitting, the shallow emission and lower signal-to-noise result in significant variation in emission spectrum between each sample. 
However, the calculated uncertainty does not reflect this specific case.
In the mean spin temperature method we limit the analysis to a window where $T_\text{b} \ge 3$K. 
For this spectrum, this results in a very narrow velocity window being analysed.
This window is imposed to avoid the calculation being dominated by noise in the emission and absorption spectra. 
Despite the narrow window, the use of bootstrapping allows the cold gas characteristics and their uncertainty to be explored. 
For this example, this results in a large uncertainty on the $f_\text{CNM}$ value.
For these types of spectra, the higher uncertainty reported by the mean spin temperature method better reflects the low signal-to-noise and varied emission nearby the source.

The other environment in which analysis is difficult is at low Galactic latitudes and in complex environment such as within the denser parts of the SMC. 
\cite{murray_2017} discussed how Gaussian fits can have a completeness as low as 29\% for low Galactic latitudes ($0 < |b| < 50\degr$) and attributed this to the presence of line blending and turbulence in the emission and absorption spectra.
As demonstrated in \cite{dempsey_2022}, the mean spin temperature method still works reliably in the presence of complex emission and absorption structure, thus providing access to cold gas characteristics.

\section{Summary and Conclusions}
\label{sec:conclusion}

In this paper we have presented the GASKAP-HI pilot observations for the Magellanic System and the dense coverage of absorption sight-lines across the field.
In all we have examined 3219 sight-lines for absorption and found potential absorption by cold \hi\ in 344 of these sight-lines.
The majority of these are in the LMC and SMC.
We have described the stacking method that was used to combine multiple observations to produce this absorption data set.
This method will be used for combining multiple observations or groups of observations for maximum sensitivity during the full GASKAP-HI survey.

Examining the cold gas in the Magellanic Bridge in detail using Gaussian decomposition, we have been able to fit and analyse 16 of 19 sight-lines with cold gas detections.
These detections follow the Young Main Sequence stars in the Bridge.
The majority of these cold gas detections are in the denser gas of the Bridge, closer to the SMC, consistent with the cold gas being either initially pulled from the SMC as the Bridge was being formed, or formed as the warm gas was pulled from the SMC.
In contrast, the cold gas fraction varies greatly across the Bridge and there is no discernible trend.
We examine one case where it seems likely that the cold gas has been created in-situ as part of a star formation cycle. 
We speculate that similar creation of cold gas by star formation and local turbulence may explain the variability of cold gas fraction across our sample.
Overall, we find a $\overline{f_\text{CNM}} = 0.12 \pm 0.08$ across the Bridge.

In Appendix \ref{sec:fourier} we have presented a map of the lower limit of cold gas column density in the Magellanic system based on Fourier transformation of the GASKAP-HI pilot spectral cubes.
This view of the large scale distribution of the cold gas in the system holds up well against the measurements of cold gas in absorption, and provides a reliable guide as to where cold gas may be found in the field.

Future deep observations of the SMC, LMC and Bridge will provide an opportunity to more closely examine the cold \hi\ gas in this region.

\section*{Acknowledgments}
This research has made use of the NASA Astrophysics Data System. 
This scientific work uses data obtained from Inyarrimanha Ilgari Bundara, the CSIRO Murchison Radio-astronomy Observatory. 
We acknowledge the Wajarri Yamaji People as the Traditional Owners and native title holders of the Observatory site. 
CSIRO’s ASKAP radio telescope is part of the Australia Telescope National Facility (https://ror.org/05qajvd42). 
Operation of ASKAP is funded by the Australian Government with support from the National Collaborative Research Infrastructure Strategy. 
ASKAP uses the resources of the Pawsey Supercomputing Research Centre. Establishment of ASKAP, Inyarrimanha Ilgari Bundara, the CSIRO Murchison Radio-astronomy Observatory and the Pawsey Supercomputing Research Centre are initiatives of the Australian Government, with support from the Government of Western Australia and the Science and Industry Endowment Fund.
This work was supported by resources provided by the Pawsey Supercomputing Research Centre’s Setonix Supercomputer (https://doi.org/10.48569/18sb-8s43), with funding from the Australia
n Government and the Government of Western Australia, under project allocation JA3.
This research was partially funded by the Australian Government through an Australian Research Council Australian Laureate Fellowship (project number FL210100039) to NMc-G.
This paper includes archived data obtained through the CSIRO ASKAP Science Data Archive, CASDA (http://data.csiro.au).
We thank the anonymous referee whose comments and suggestions have improved this manuscript.

\section*{Data Availability}

The observational data described in this paper are available under project AS108 in the CSIRO Astronomy Science Data Archive at \url{https://data.csiro.au/domain/casda}.
The absorption spectra, catalogues and derived data produced in this study are published as \cite{Dempsey2026}.

\bibliographystyle{mnras}
\bibliography{references} 

\appendix

\section{Stacking Absorption Cubelets}
\label{sec:stacking-app}

\begin{figure}
  \centering
  \includegraphics[width=\linewidth]{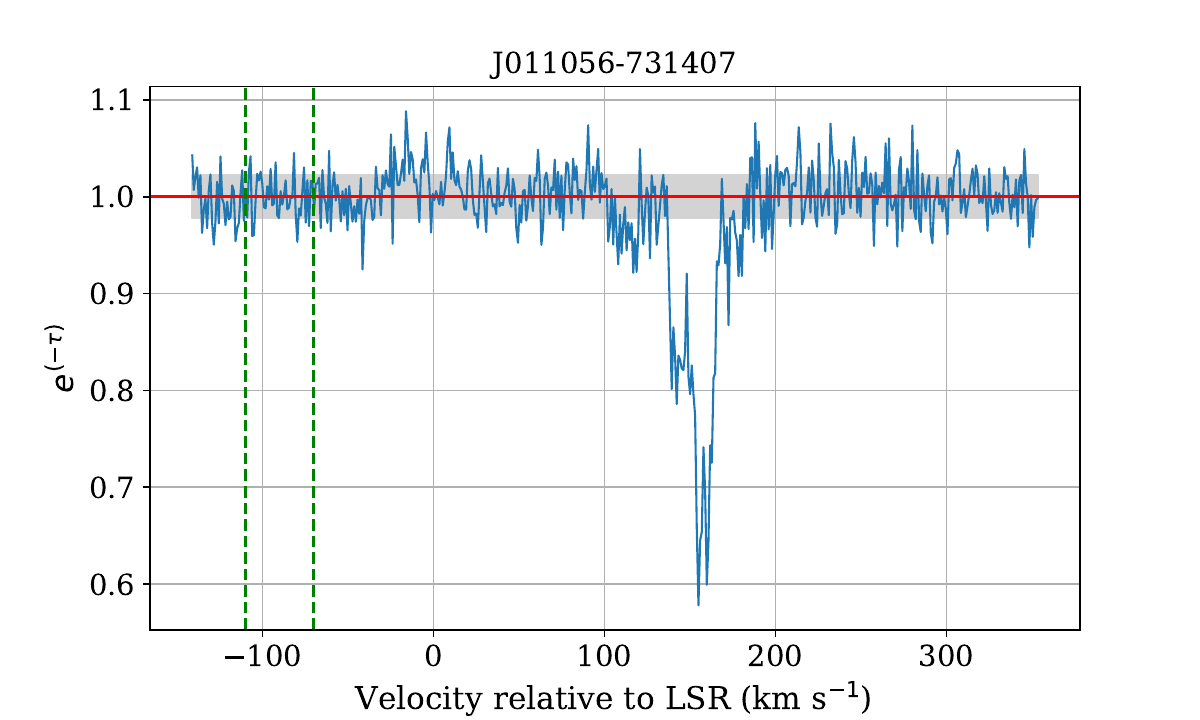}
  \includegraphics[width=\linewidth]{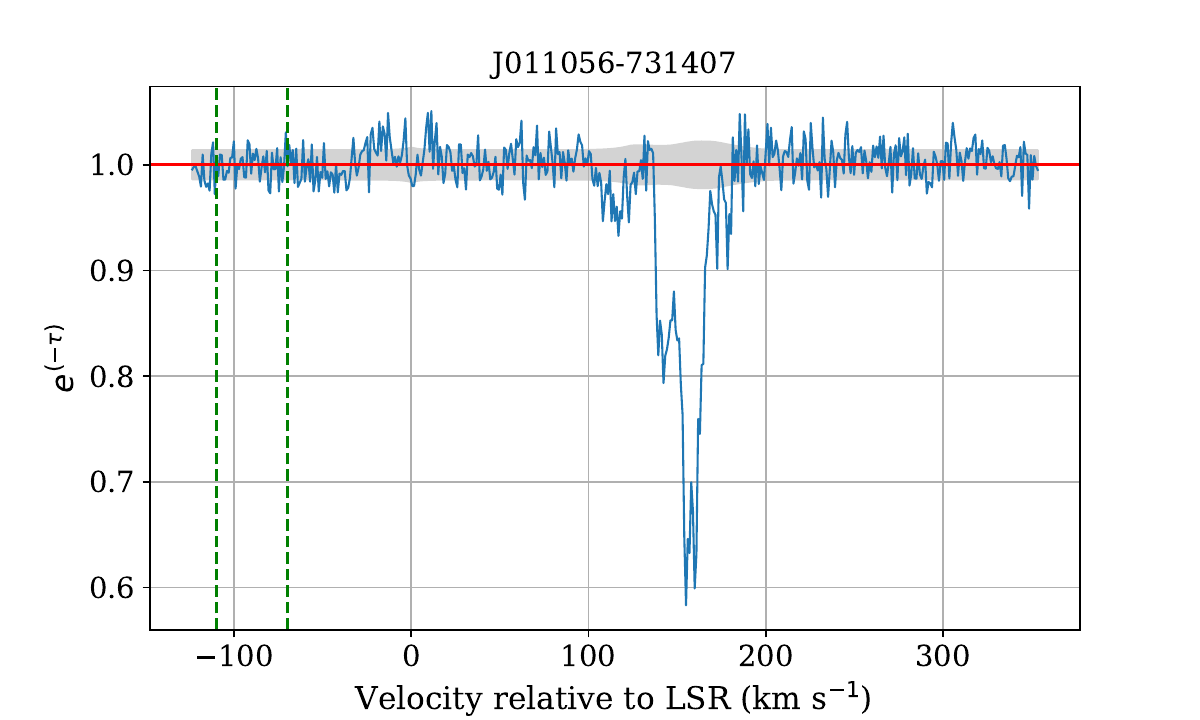}
  \caption{Comparison of spectra before and after stacking. The top panel shows the spectrum before stacking (from observation 10944) with a $\sigma_\text{cont} = 0.022$. The bottom panel shows the result after stacking, giving a reduced noise level of $\sigma_\text{cont} = 0.013$}
  \label{fig:stacking}
\end{figure}

Where a sight-line has been observed more than once, we use stacking to achieve an optimal signal-to-noise spectrum.
Although this process is described in Section \ref{sec:data}, we present more details here as well as some examples of spectra before and after stacking in Figure \ref{fig:stacking}. 

\begin{figure}
  \centering
  \includegraphics[width=\linewidth]{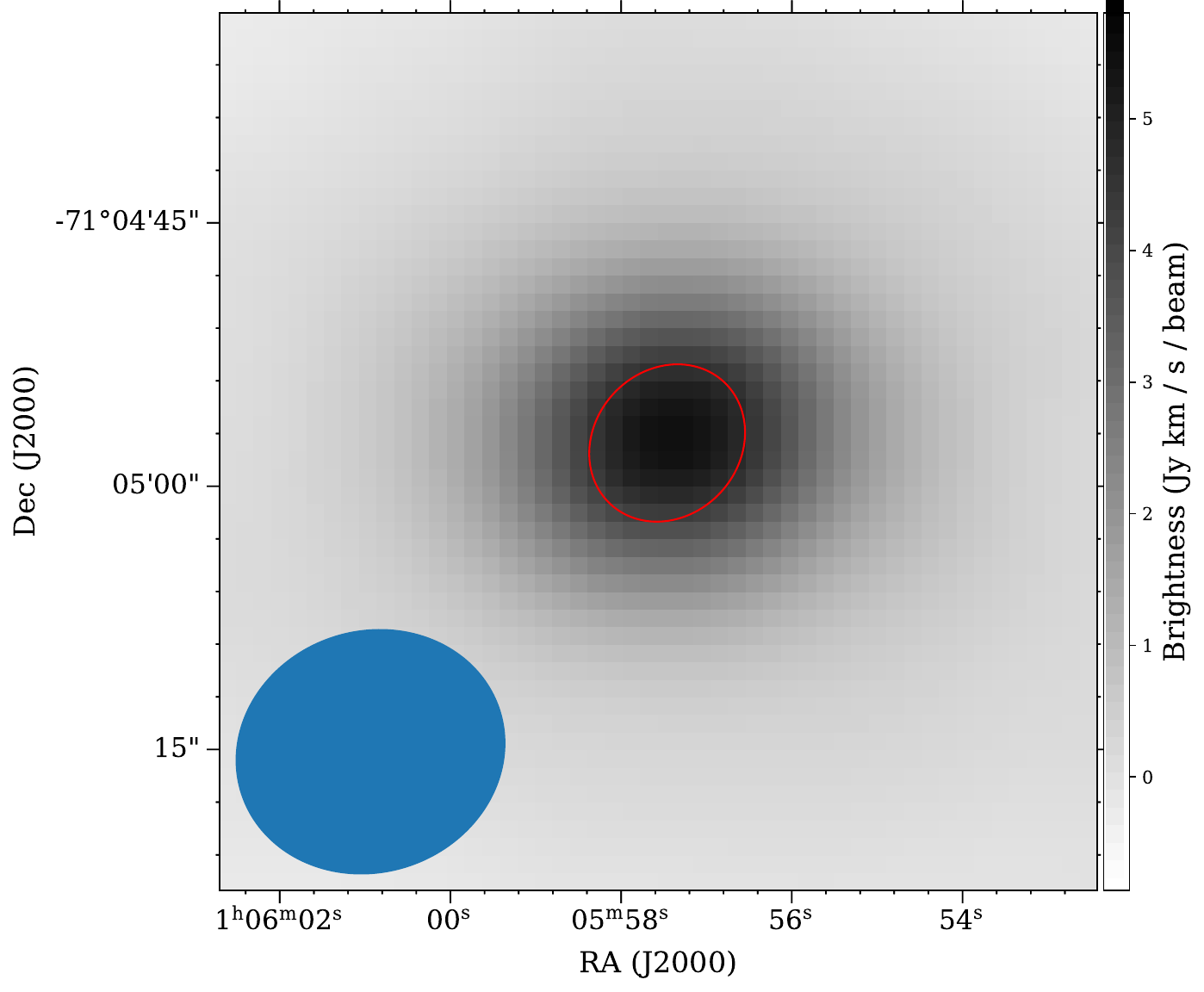}
  \caption{Example cubelet moment-0 map for source J010557$-$710457 from observation 30665. The blue filled ellipse shows the synthesised beam size and the red ellipse indicates the source extent as identified from the continuum map.}
  \label{fig:mom0-map}
\end{figure}

As noted in Section \ref{sec:data} we have chosen to stack the per-source cubelets in the image domain.
In radio interferometry, stacking is more often performed in the Fourier domain by joining visibility files and cleaning the larger single data set.
This is often done to improve the cleaning response on low surface brightness diffuse emission.
In contrast, we have excluded shorter base lines to filter out such diffuse emission and use limited cleaning cycles as we are focussed on the compact continuum sources.
The resulting cubelets are generally simple single source images such as is shown in Figure \ref{fig:mom0-map}.
Stacking these pilot survey cubelets in the image domain allows us to validate the approach we will need to take in the full survey when we do not have access to all visibilities due to data storage limitations.

As a test of the effectiveness of the image domain stacking approach we have processed the pilot phase II observations of the SMC using both image-domain and Fourier-domain stacking and compared the results.
We found that for 215 of the 386 spectra ($56\%$) the noise from both methods was within 5\% of each other. For a further 95 spectra ($25\%$) we found the image-domain stacked spectra had $<10$\% higher noise than for the Fourier-domain stacked spectra.

Across the 868 spectra stacked in processing pilot phase I and II GASKAP-HI data, we obtain a median improvement of 15\%. In only 67 of the 868 (or 8\%) sightlines did we find stacking resulted in marginally increased noise and thus we defaulted to the best spectrum instead.

\section{Comparison with the M24 method}
\label{sec:fourier}
\subsection{A lower limit on the column density of cold gas}
\begin{figure*}
  \centering
  \includegraphics[width=\linewidth]{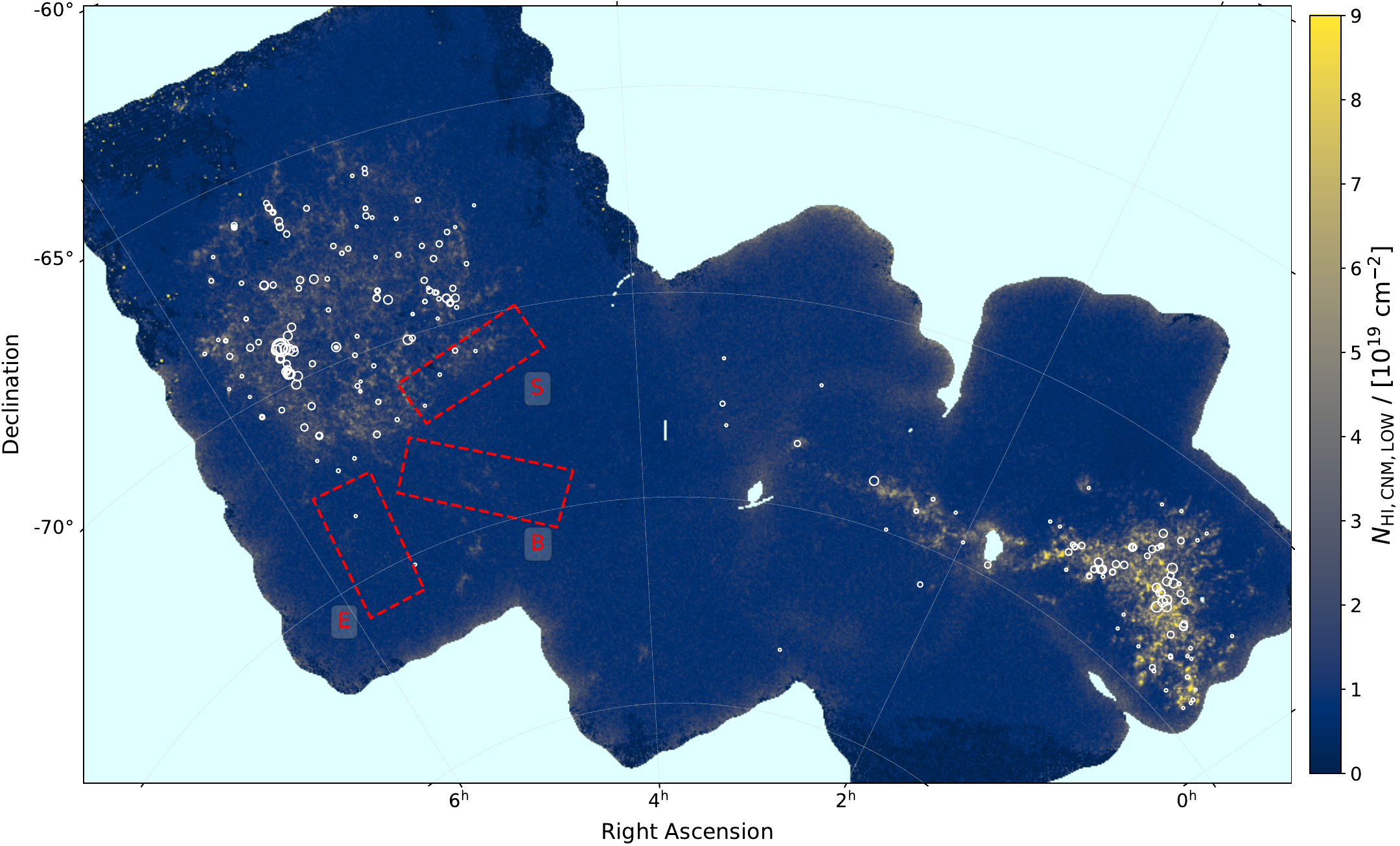}
  \caption{Map of the lower limit on the Magellanic system cold gas column density $N_\text{CNM,low}$ in units of 10$^{19}$\,cm$^{-2}$ of the entire GASKAP-HI Pilot I and II survey dataset showing column densities of cold gas detected in absorption.
  The candidate detections with significance $\geq 4\sigma$ are shown with white circles and the size of each circle encodes the column density of cold gas.
  The pale blue holes and arcs in the image are regions which have been masked out as the primary beam power was less than 25\%.
  The LMC E, B and S arms \citep{staveley-smith_2003} are outlined in red from left to right.
  }
  \label{fig:NHI_CNM_FFT_Abs_Comparison}
\end{figure*}
We made use of the method developed in \citetalias{marchal_2024} that quantifies a lower limit of the cold gas mass fraction from emission spectra only, using the discrete FT of the brightness temperature $T_b(v)$: 
\begin{equation}
    \hat T_b(k_v) = \sum_v T_b(v) \, \exp \left( -2\,\pi\,j\,v\,k_v\right) \, dv \, ,
\end{equation}
where $j$ is the imaginary unit, and $k_v$ denotes the Fourier conjugate of the velocity $v$.
$\hat T_b(k_v)$ provides a frequency representation of the brightness temperature spectrum and is informative of the presence of narrow features in $T_b(v)$. 
In other words, a $\hat T_b(k_v)$ exhibiting high power at small velocity scales (i.e., large $k_v$) indicates the presence of cold gas in the emission spectrum, though it provides only a lower limit on its absolute value. 
We refer the reader to \citetalias{marchal_2024} for an in depth discussion on the origin of this lower limit, mainly due to interference pattern seen in the FT as well as an opacity effect.

We define the cold gas mass fraction as 
\begin{align}
   f^{k_{v,\rm lim}}_{\rm low} = \hat T_b(k_{v,\rm lim}) / \hat T_b(0) \, ,    \label{eq:mass_fraction}
\end{align}
where $\hat T_b(0)$ is the value of the FT at $k_v=0$ and is proportional to the total column density of the line in the optically thin limit. 
Note that this differs slightly from Equation~9 of \citetalias{marchal_2024}, which considers the highest value to the right of the selected limit $k_{v,\rm lim}$ (a prescription suited for high signal-to-noise spectra). 
In the data analysed here, however, noise in the emission data is significant, and high noise levels tend to dominate the FT at large $k_v$. 
By selecting the value at $k_{v,\rm lim}$, we obtain a more robust lower limit, reducing (though not eliminating) the influence of noise in the FT map.
Here, we adopted the recommended threshold $k_{v,\rm lim}=0.12$\,(\kms)$^{-1}$, which ensures the suppression of 95\% of any Gaussian components with a velocity dispersion greater than 3\,\kms.

The lower limit on CNM density, $N_\text{CNM,low}$ is shown in Figure~\ref{fig:NHI_CNM_FFT_Abs_Comparison} (hereafter the FT map) in a combined mosaic map of all the GASKAP-HI fields tabulated in Table~\ref{tab:observations}. . 
The candidate detections with significance $\geq 4\sigma$ are shown with white circles and the size of each circle encodes the column density computed in the optically thin limit.
The FT map shows abundant small-scale features, especially in the LMC. 
While in principle this invites a comparison of the properties of cold gas between SMC, LMC, and MB, the fact that the FT map only provides a lower limit complicates a morphological analysis. 
The apparent `speckled' structure of the gas in the LMC could be due to more complex line shapes which can produce interferences that do not reflect physical structures. 
In the SMC, though it appears more coherently structured, the known multiple-peak profiles \citep{murray_2024} likely also modulates the FT map through similar interferences. 
Extracting reliable information about the absolute amount and/or morphology of cold gas using the FT map pushes the intrinsic limits of a lower-limit estimator. 
Here, the strength of the FT method lies in capturing the large scale distribution of cold gas rather than its fine-scale morphology and we find that on those scales, there visually is a strong correlation between the positions of detections and the positions where cold gas is predicted by the FT.
Together, they provide a consistent and comprehensive large scale picture of where cold gas exists in the MCs and the MB.

As inferred in Section~\ref{sec:bridge-abs} from the absorption catalogue alone, cold gas is predominantly found on the western side of the MB, where it connects to the SMC and where the total column density of \hi\ is highest.
As noted previously, this is also where most of the YMS stars in the Bridge are found \citep{Belokurov_2017}.
Following the nomenclature of \citet[][]{staveley-smith_2003}, both the FT map and the absorption catalogue reveal the presence of cold gas in the LMC Arm 'S', which borders the main body of the LMC \citep[][]{staveley-smith_2003}, as well as, though more sparsely, in Arm 'E', which extends in the direction of the Leading Arm.
Conversely, unlike the detected absorption sources\footnote{%
It should be noted, however, that the conservative cut for candidate detection with a significance $\geq 4\sigma$ may be responsible for this difference.}, the FT map shows isolated cold gas clumps associated with Arm 'B', which links to the Magellanic Bridge.
The arms are annotated in Figure~\ref{fig:NHI_CNM_FFT_Abs_Comparison} with red boxes.
A more distant cloud of cold gas is detected in the direction of Arm 'B' at $\alpha = 04^\mathrm{h}19^\mathrm{m}20\fs55,\ \delta = -74^\circ34^\prime20\farcs98$
, beyond which no obvious continuation of cold gas toward the Bridge is observed in the FT map.
Although no sharp boundary distinguishes LMC-originating from SMC-originating gas along the Magellanic Bridge, this apparent break in cold gas seen in the FT map may offer clues about their respective origins, and/or local ISM conditions driving the condensation or recycling of cold gas. 

\subsection{Quantitative comparison with cold \hi\ seen in absorption}

\begin{figure}
  \centering
  \includegraphics[width=\linewidth]{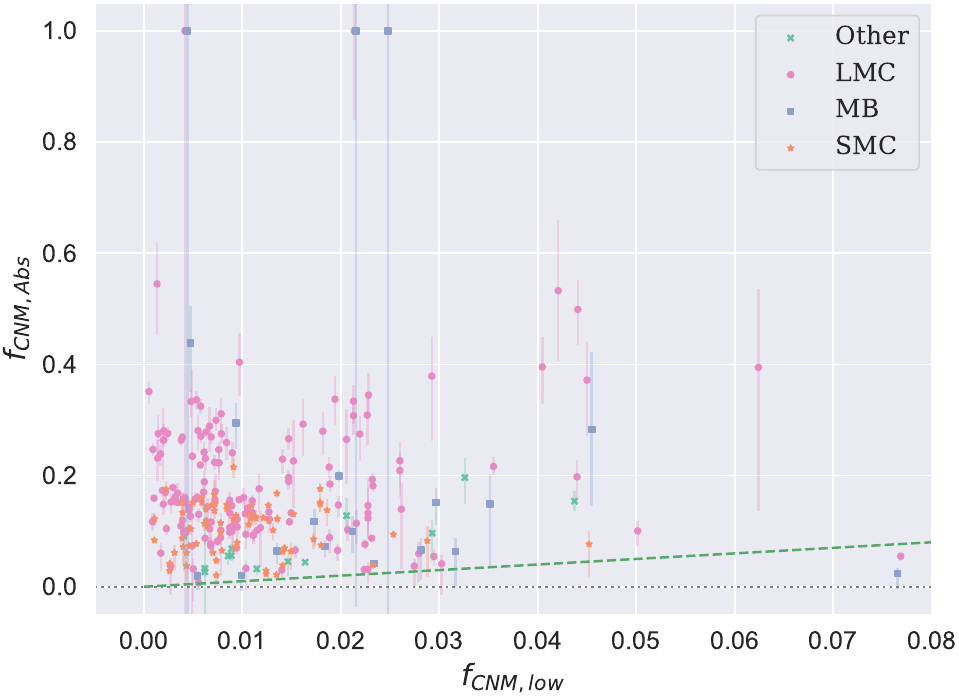}
  \caption{Comparison of the fraction of cold gas found in the absorption spectra ($f_\textrm{CNM,Abs}$) vs the lower limit on cold gas fraction found in the FT method ($f_\textrm{CNM,low}$) for each detection.
  The structure a source is associated with is indicated by its colour and symbol.
  The 1$\sigma$ uncertainty range in the CNM measurement is shown as a vertical line for each detection.
  The green dashed line shows equality between the two axes.
  Three LMC sources with $f_\textrm{CNM,low} > 0.08$ are excluded from this plot for clarity. 
  One of these sits below the green line, i.e. $f_\textrm{CNM,low} > f_\textrm{CNM,Abs}$.
  }
  \label{fig:fcnm_comparison}
\end{figure}

\begin{figure}
  \centering
  \includegraphics[width=\linewidth]{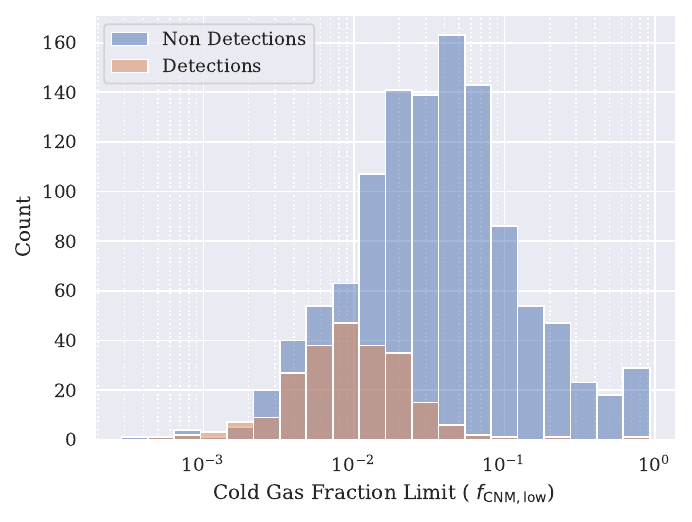}
  \caption{Comparison of the distribution of spectra with and without candidate absorption  detections compared with the lower limit on cold gas fraction found in the FT method ($f_\textrm{CNM,low}$) for those lines of sight.
  The blue bars show the count of spectra without detected absorption features in each column density bin, while the orange bars show the count of spectra with absorption features in those bins.}
  \label{fig:fcnm_det_hist}
\end{figure}

Next, we numerically compare the map of the lower limit of cold \hi\ through the FT approach with the direct measurements of cold \hi\ through absorption. 
In Figure \ref{fig:fcnm_comparison} we compare the lower limit with the measured cold \hi\ fraction.
We can see that in all cases, apart from three, the lower limit is below the measured cold gas fraction.
However, it is also apparent that in most cases the limit is conservative and there is no discernible trend between the lower limit and the measured cold \hi\ fraction other than the fraction being above the limit.
\cite{murray_2017} note that line-blending and turbulence in dense emission environments can make the different components difficult to disentangle.
This is likely a factor in the conservative limit on cold gas seen in our denser environments such as the SMC.
Looking at the cold gas fractions, most very high CNM fractions measured in absorption have a high uncertainty, likely reflecting a complex emission environment.
The SMC sightlines have a low CNM fraction and lower uncertainty than other regions. 
The lower uncertainty reflects the longer dwell time on this field.
In contrast, the LMC exhibits a wide range of measured cold gas fractions and has some of the largest CNM fraction limits.
The region also has higher uncertainty levels than the SMC, as expected with the shorter observation time.

Analysing the sight-lines where there was no significant detection of cold gas, we find that the estimated cold gas fraction from absorption is above the FT limit in the vast majority of cases. 
However, the \FCNM\ estimates have large uncertainties, particularly for sight-lines with low signal to noise ratio in the surrounding emission. 
The limit is also well below the estimate in many cases.
Comparing the $f_\textrm{CNM,low}$ of the detections and non-detections, as shown in Figure \ref{fig:fcnm_det_hist}, we find that the median limit is higher for non-detections (0.04) than it is for  detections (0.009). 
This is the opposite of what we would expect if the limit were predicting where the cold gas is present. 
Many of the non-detections with higher FT limits are on the far left and on the outer edges of the fields where there is higher noise and limited emission resulting in a quite low signal to noise in the emission data. 
In these locations, narrow line noise can be picked up as cold gas.
This is because the GASKAP-HI pilot data has a noise level (1.1K per 0.98 km s$^{-1}$ channel; \citealt{pingel_2022}) much higher than the noise level of GASS (57 mK per 1 km s$^{-1}$ channel; \citealt{kalberla_2010_gass}), the data which were used in  \citetalias{marchal_2024}.

Despite those issues with noise though, the map of cold gas density (Figure \ref{fig:NHI_CNM_FFT_Abs_Comparison}) still shows an excellent correlation to the positions of cold gas. 
The reason is that, as shown in Figure \ref{fig:detections_nh_hist}, \hi\ column density is a good predictor of the presence of cold gas. 
So when the FT limit is combined with $N_{\mathrm{H\,I}}$ to determine a minimum cold gas density, the result is a reliable guide as to where cold gas may be found in a field.
With upcoming GASKAP-HI observations featuring deep integrations over the Magellanic Clouds and surrounding regions, and thus significantly pushing down the noise, we look forward to further enhancements to this map of a lower limit of cold gas density.

\section{Magellanic Bridge Spectra Gaussian Fits}
\label{sec:bridge-spectra-plots}

Here we present plots of the results of joint Gaussian fitting of the emission and absorption spectra for each of the sightlines towards the Bridge.
In each panel of Figure \ref{fig:bridge-spectra} we show the absorption spectrum paired with the emission spectrum at the closest point to the target continuum source but not directly at the continuum source.

\begin{figure*}
\centering
  \includegraphics[width = 3.5in]{J013218-715348_em_fit_2.pdf}
  \includegraphics[width = 3.5in]{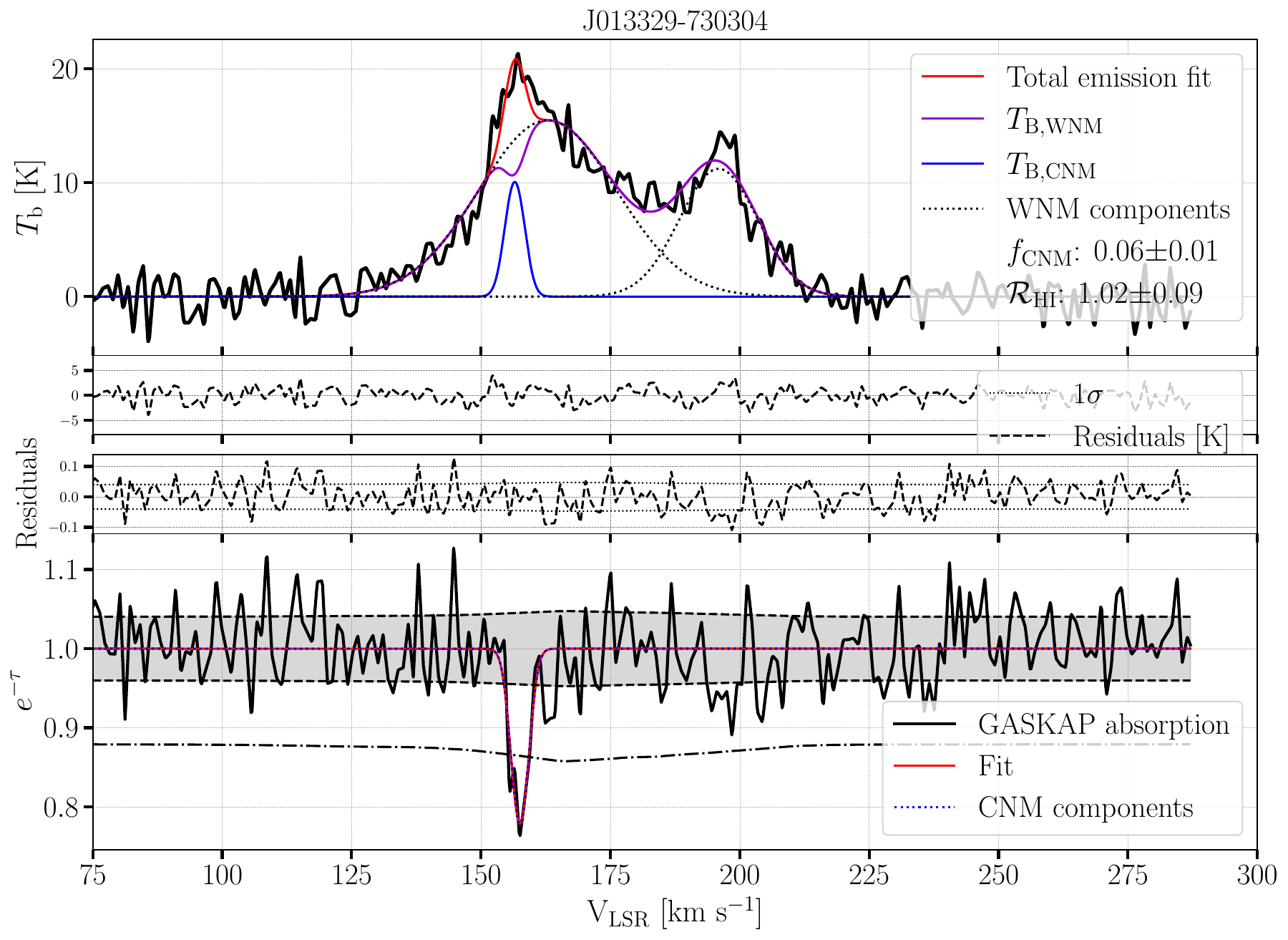}
  \includegraphics[width = 3.5in]{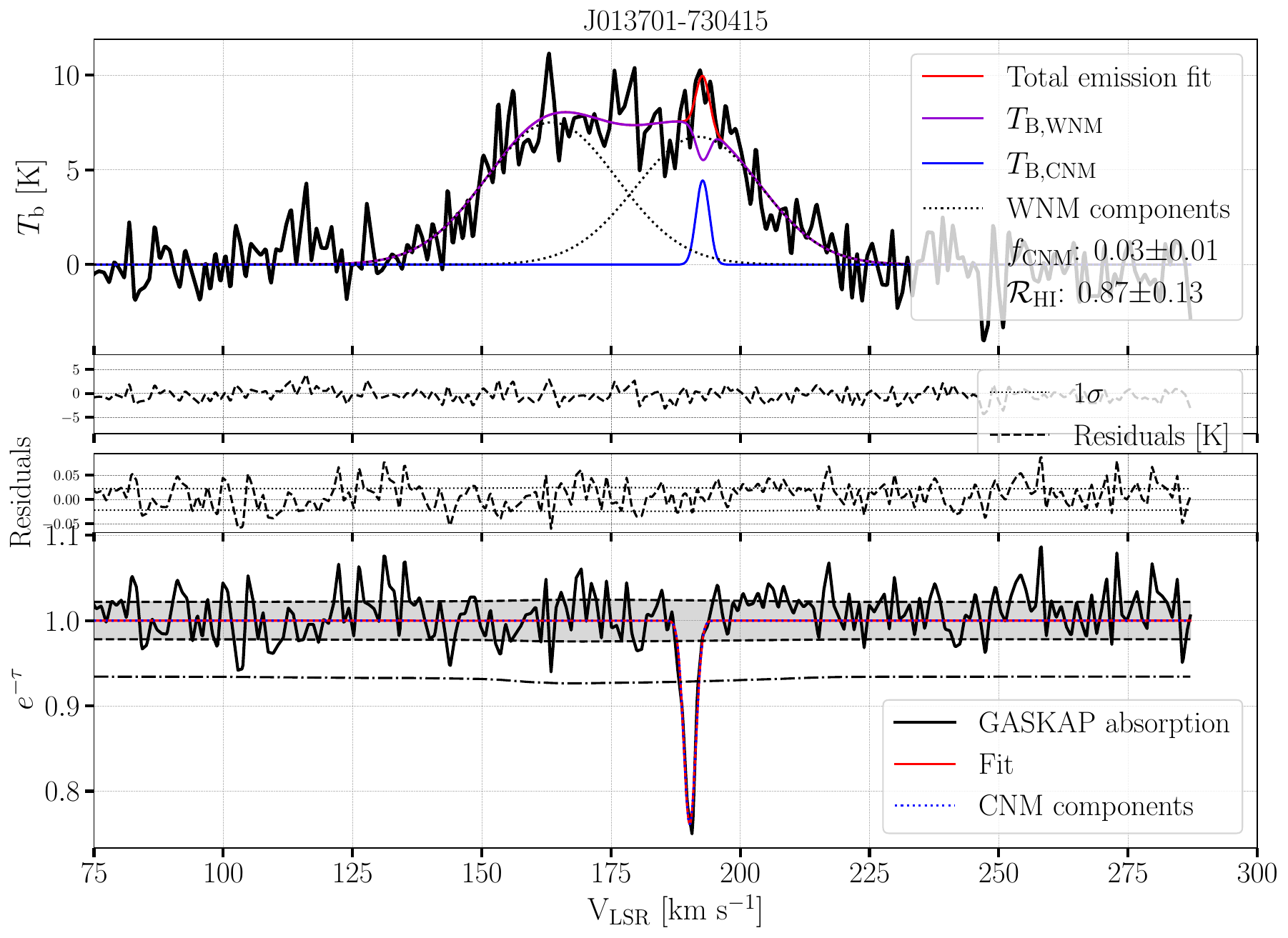}
  \includegraphics[width = 3.5in]{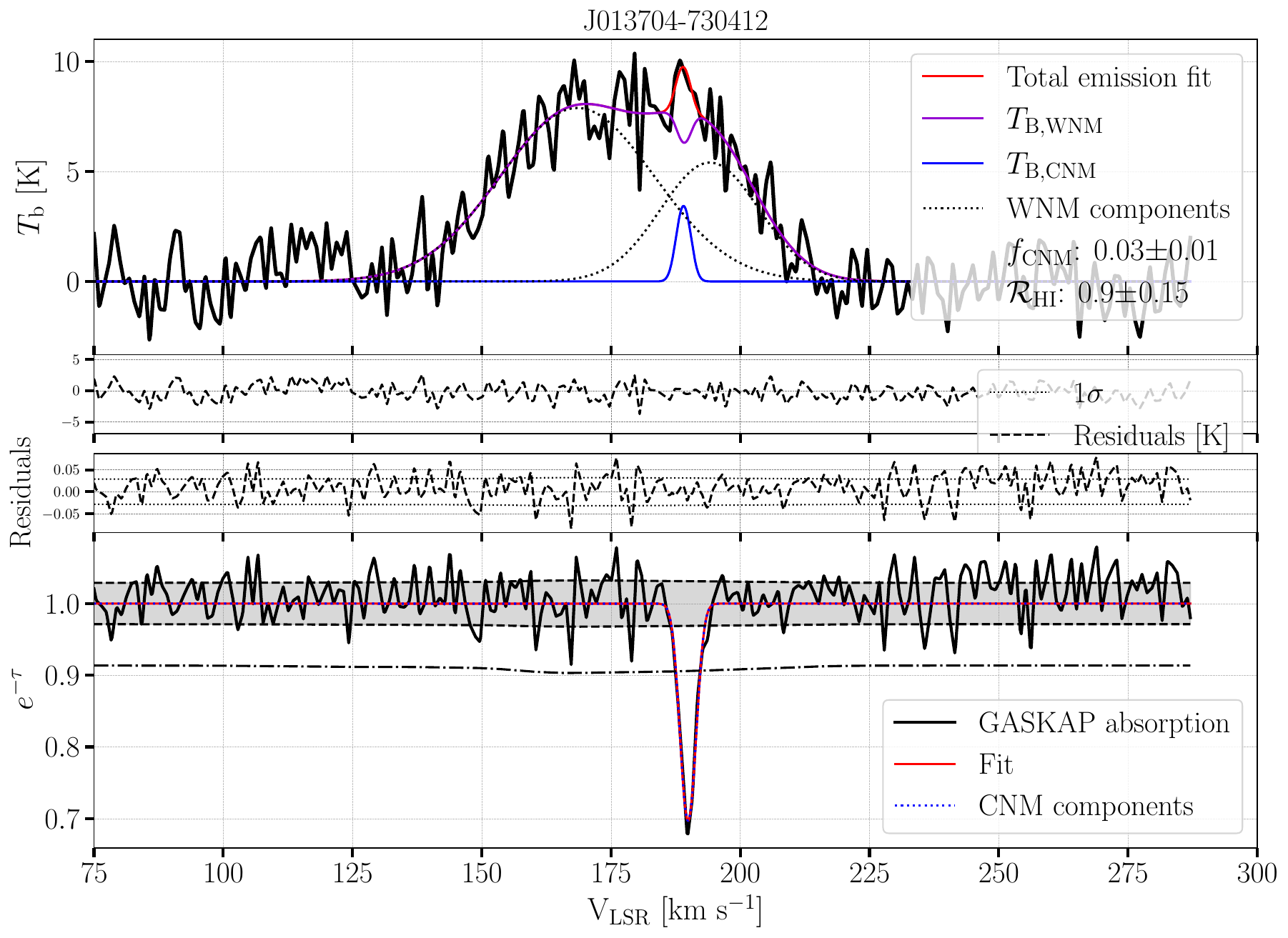}
  \includegraphics[width = 3.5in]{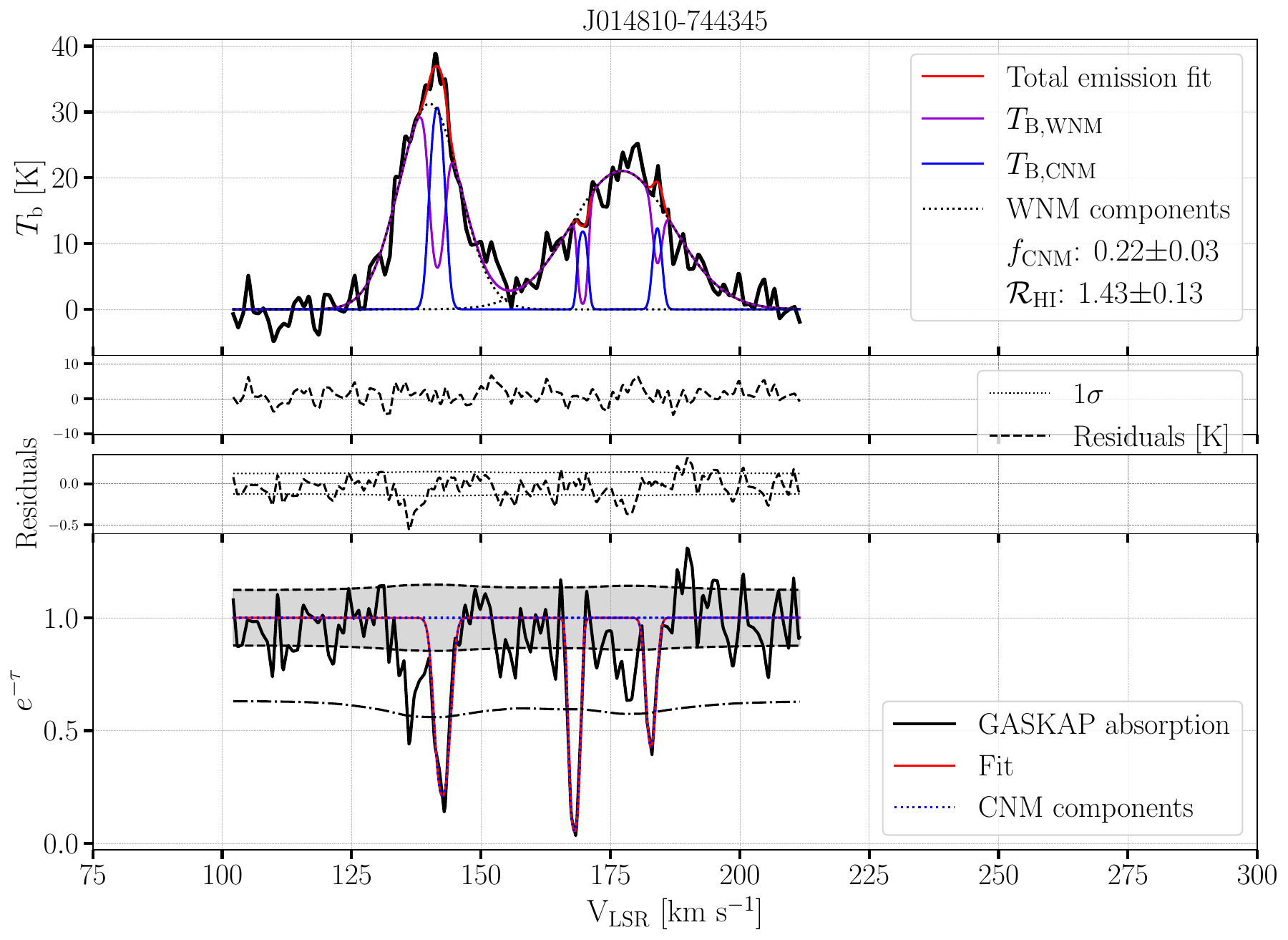}
  \includegraphics[width = 3.5in]{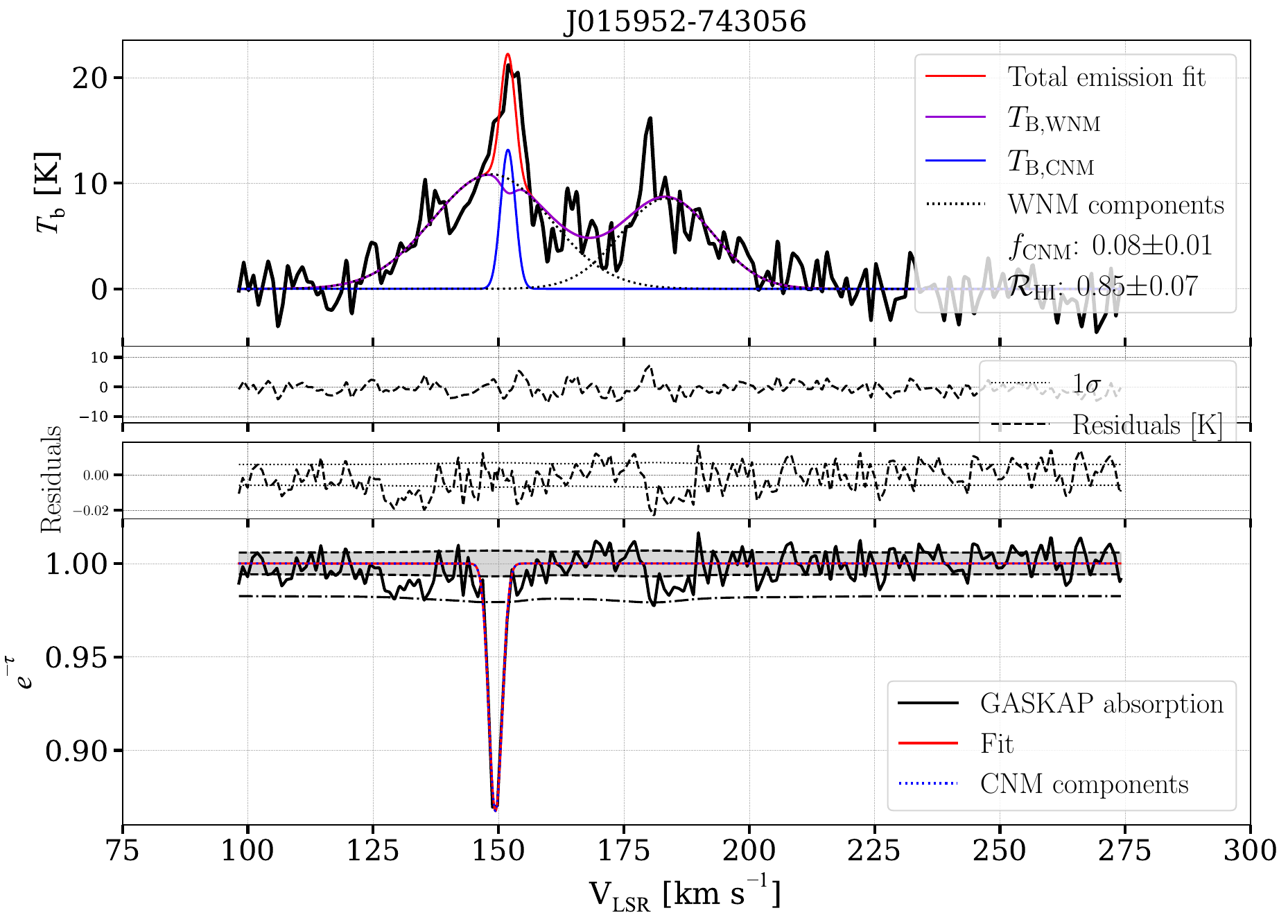}
  \caption{Gaussian decomposition of emission and absorption spectra for Magellanic Bridge sources. In each panel, the lower section shows the absorption spectrum ($e^{-\tau}$) in black with the fit shown in red. The one-sigma noise envelope is shown as gray shading and the three sigma threshold is shown as a dot-dashed line. A residual between the fit and the spectrum is shown at the top of the lower panel. The upper panel shows the brightness temperature ($T_b$) spectrum in black with the fit shown as the red line. The individual WNM components are shown as dotted lines and the CNM component is shown as a blue line. Again the residual between the fit and the emission spectrum is shown at the bottom of the top panel.}
  \label{fig:bridge-spectra}
      
\end{figure*}

\begin{figure*}
\centering
  \ContinuedFloat
  \includegraphics[width = 3.5in]{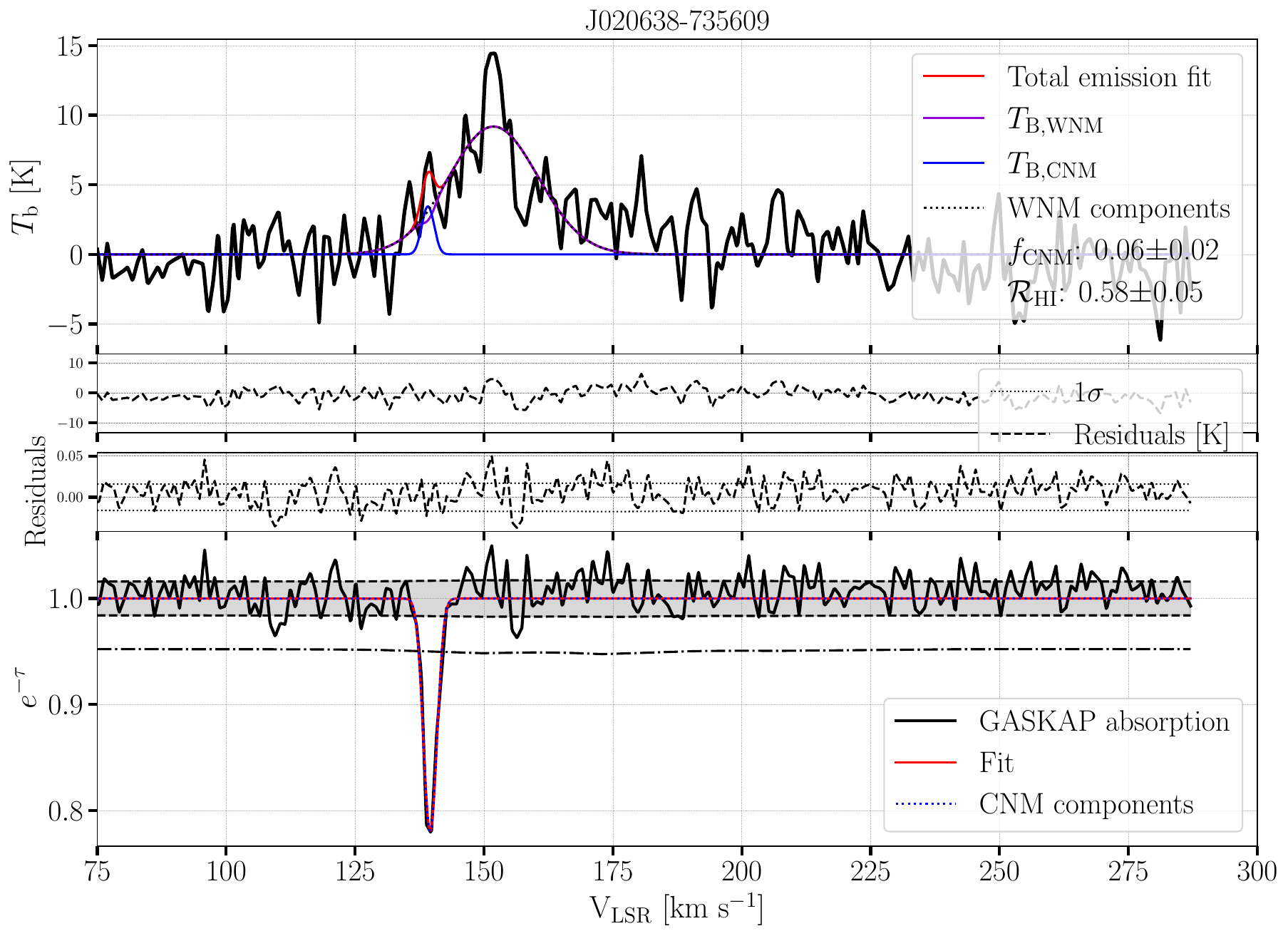}
  \includegraphics[width = 3.5in]{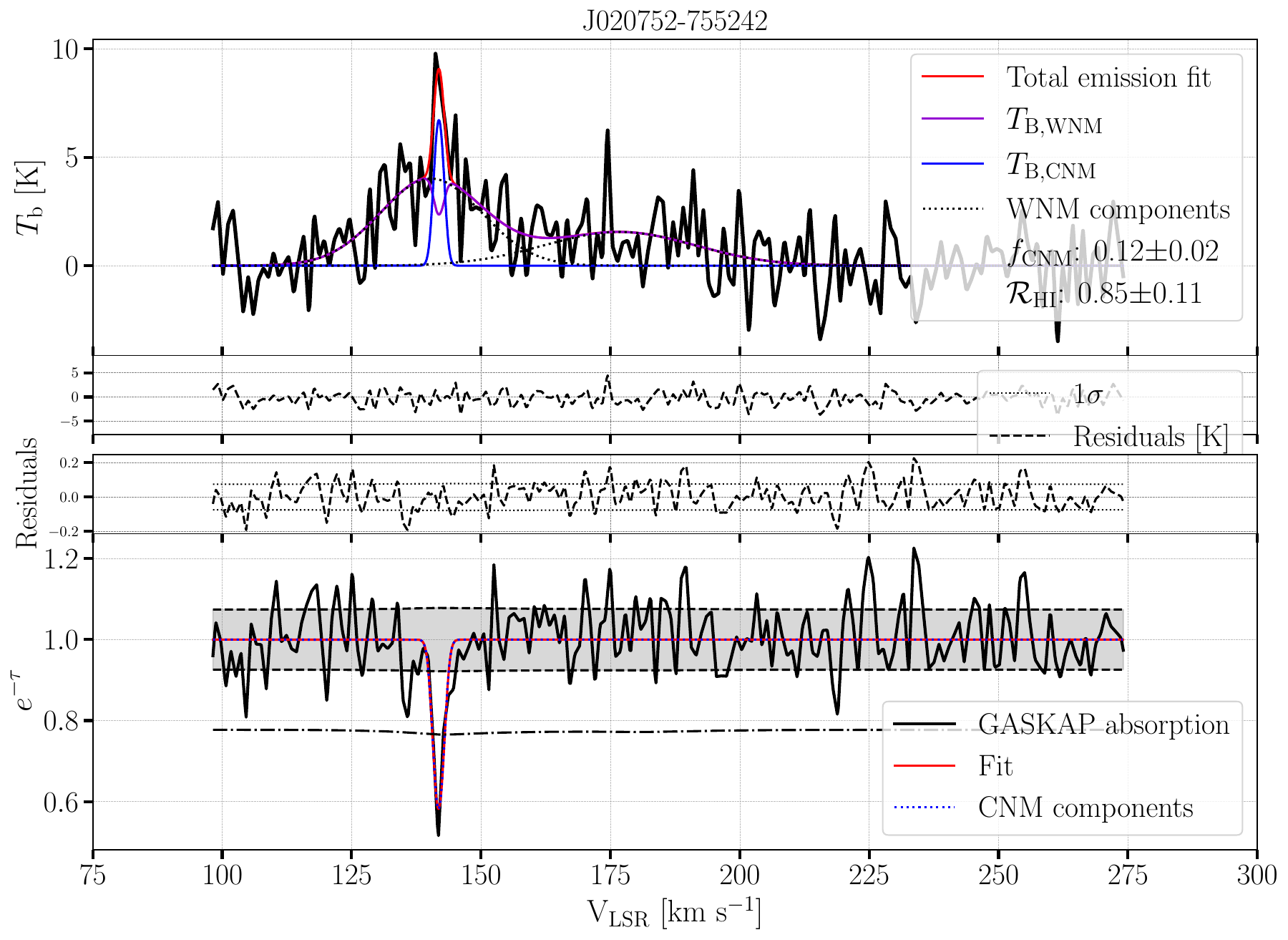}
  \includegraphics[width = 3.5in]{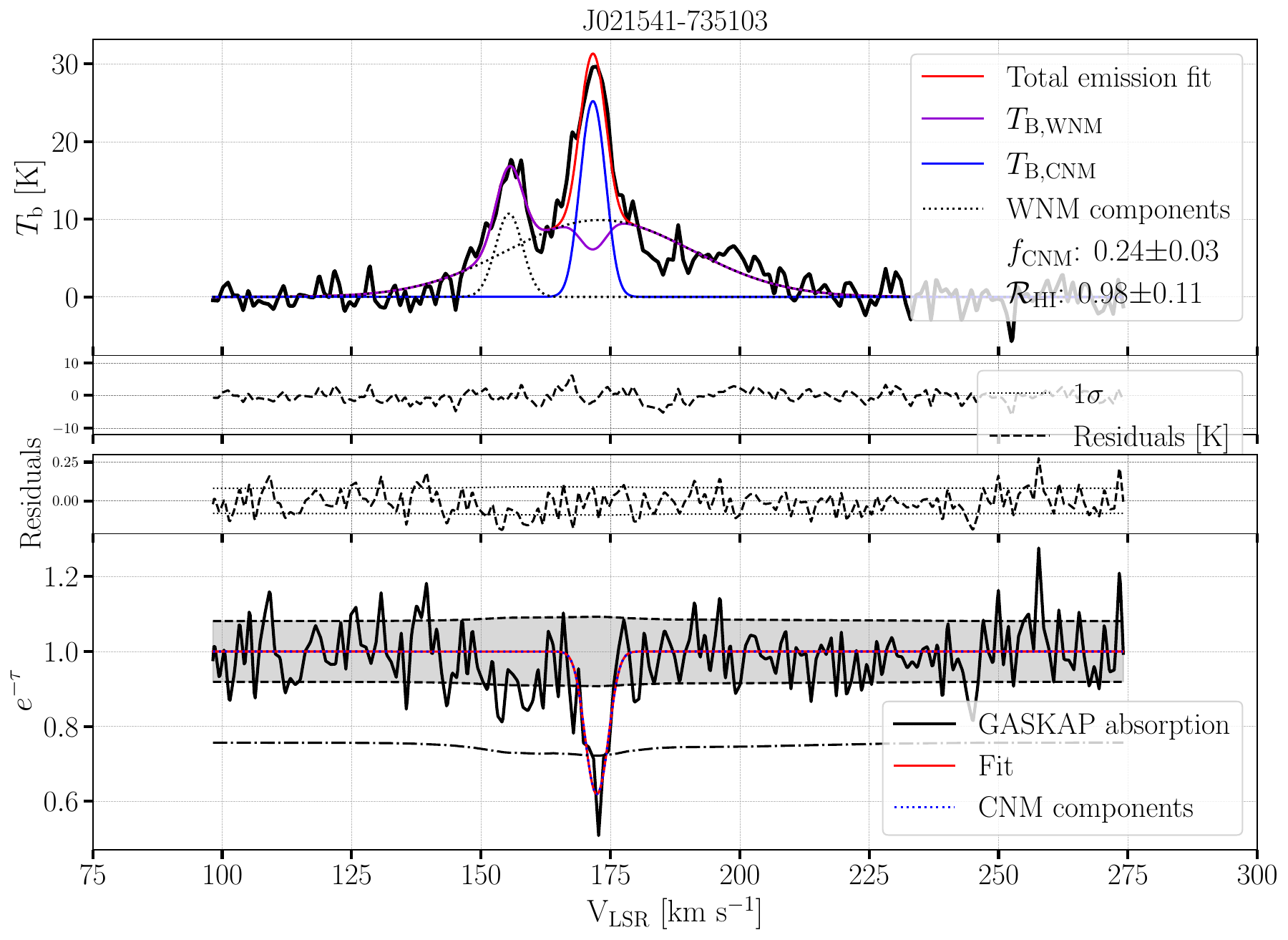}
  \includegraphics[width = 3.5in]{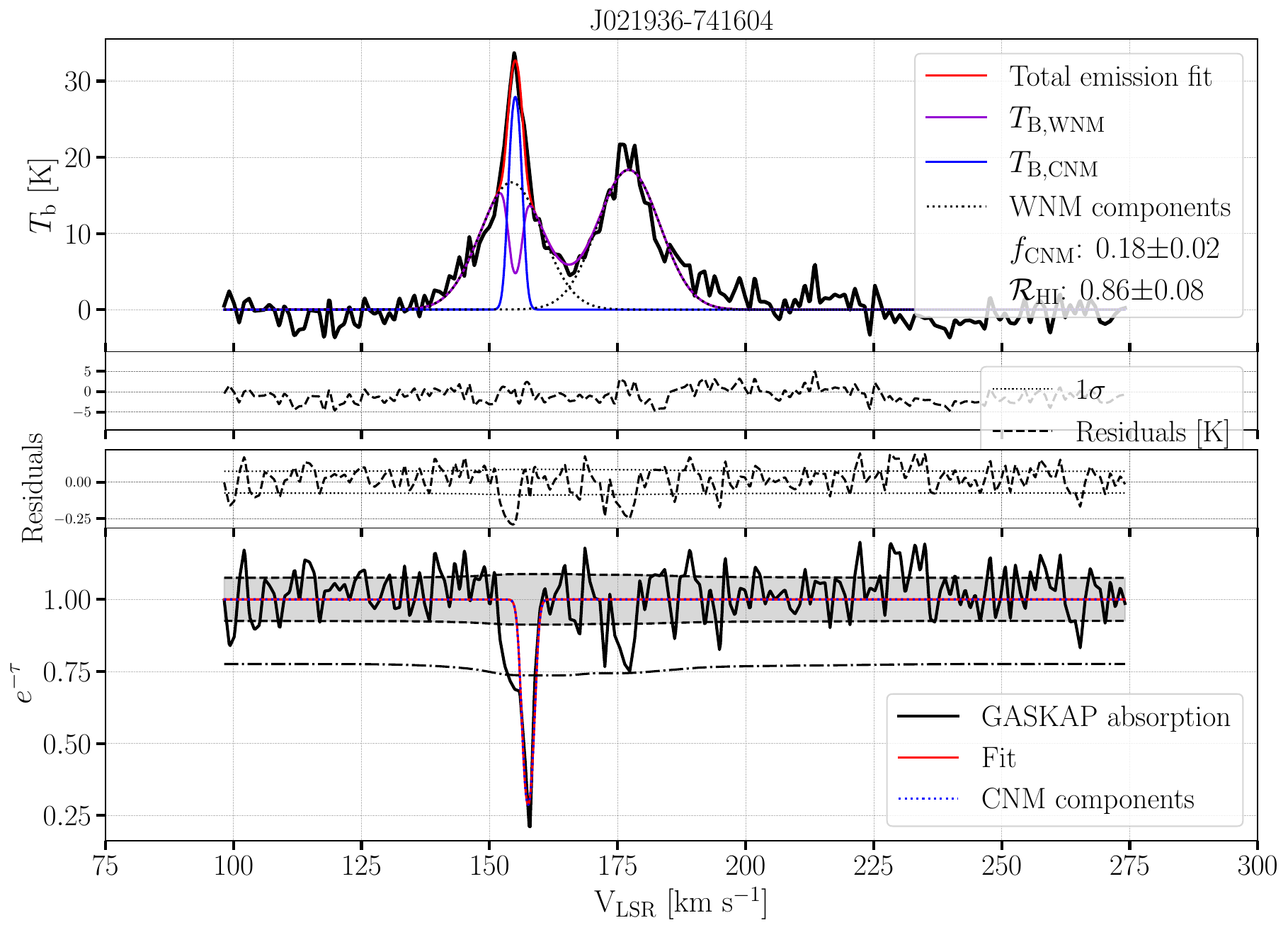}
  \includegraphics[width = 3.5in]{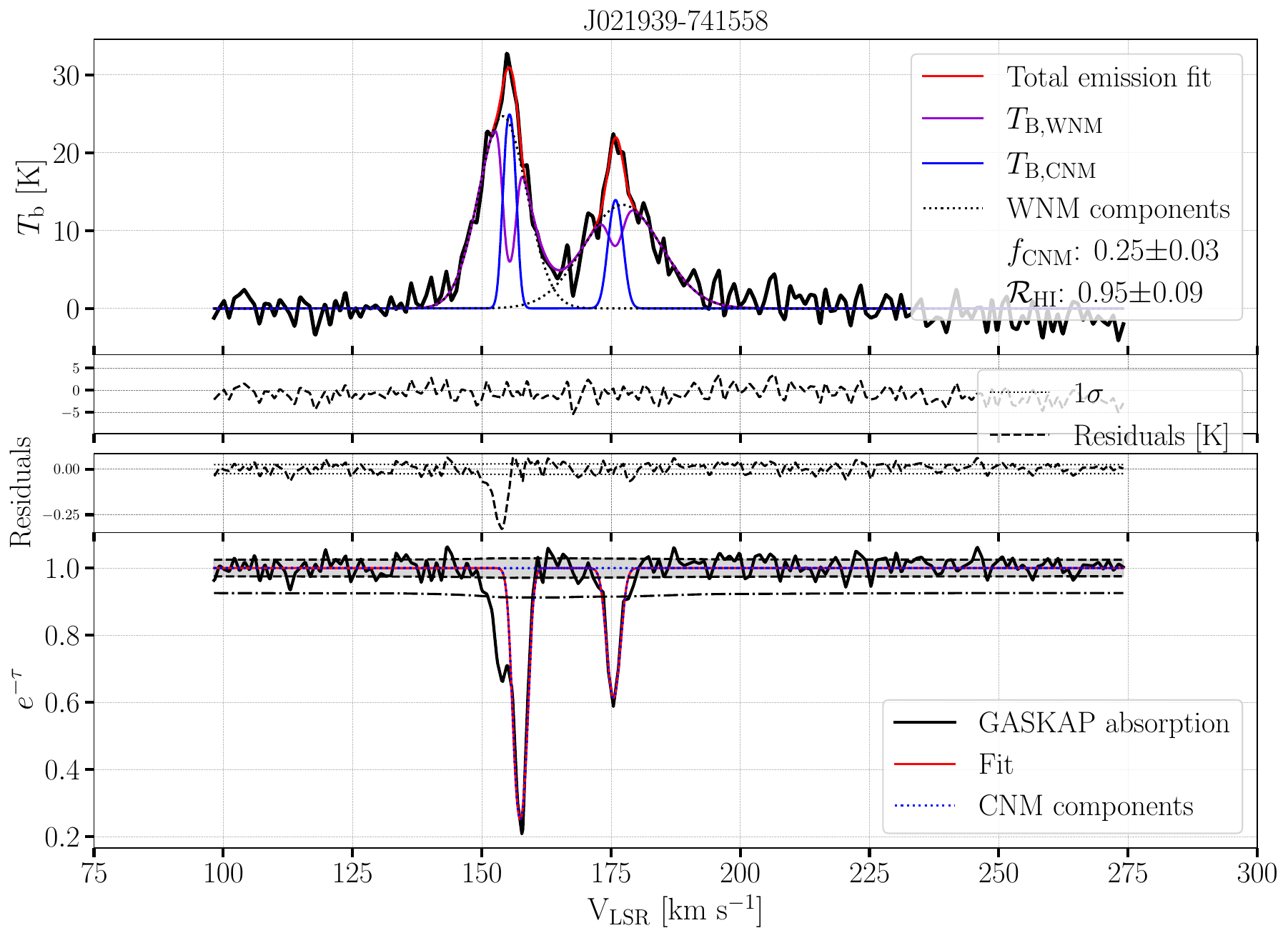}
  \includegraphics[width = 3.5in]{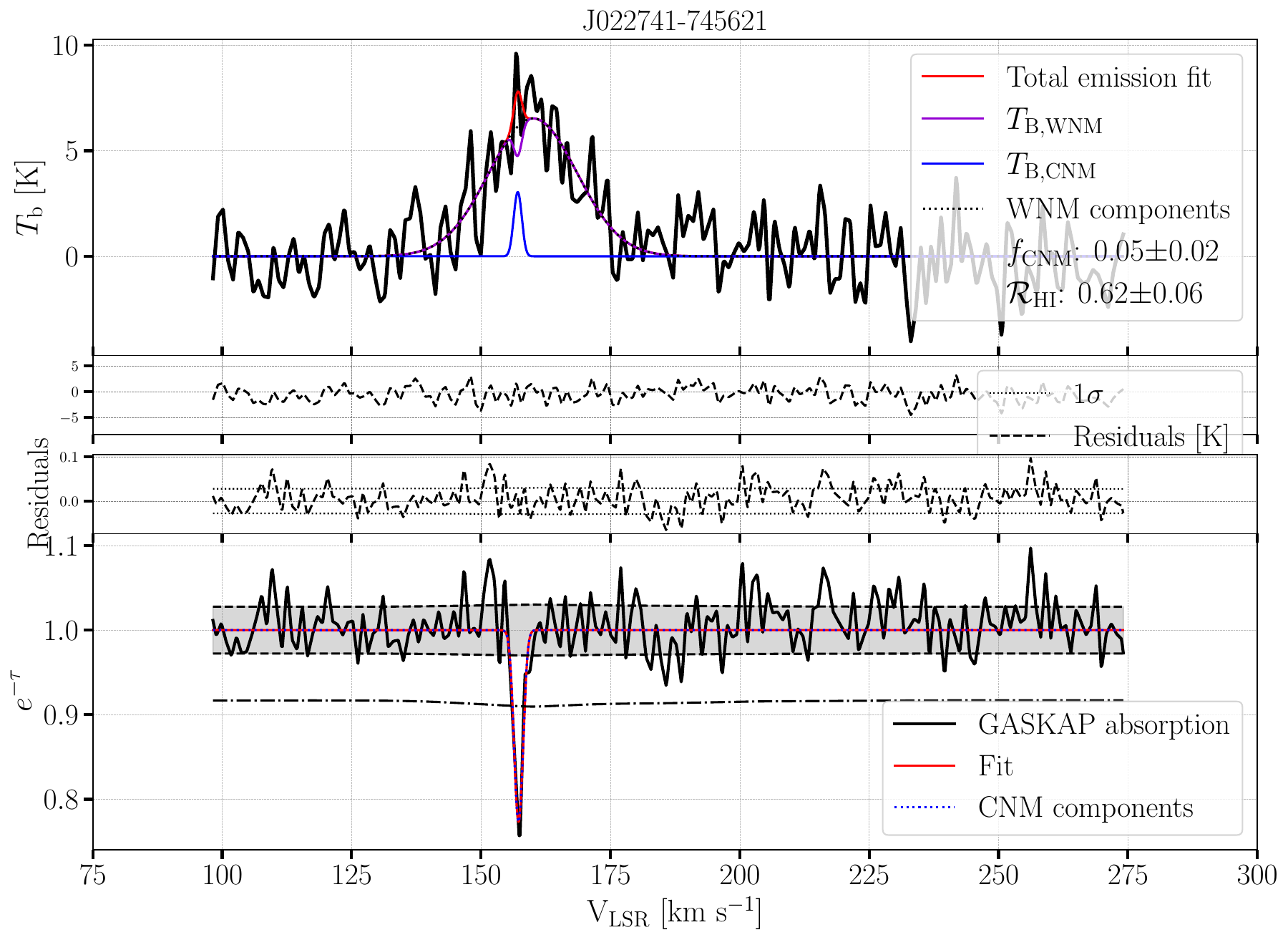}
  \caption{Gaussian decomposition of emission and absorption spectra for Magellanic Bridge sources (cont.)}
      
\end{figure*}

\begin{figure*}
\centering
  \ContinuedFloat
  \includegraphics[width = 3.5in]{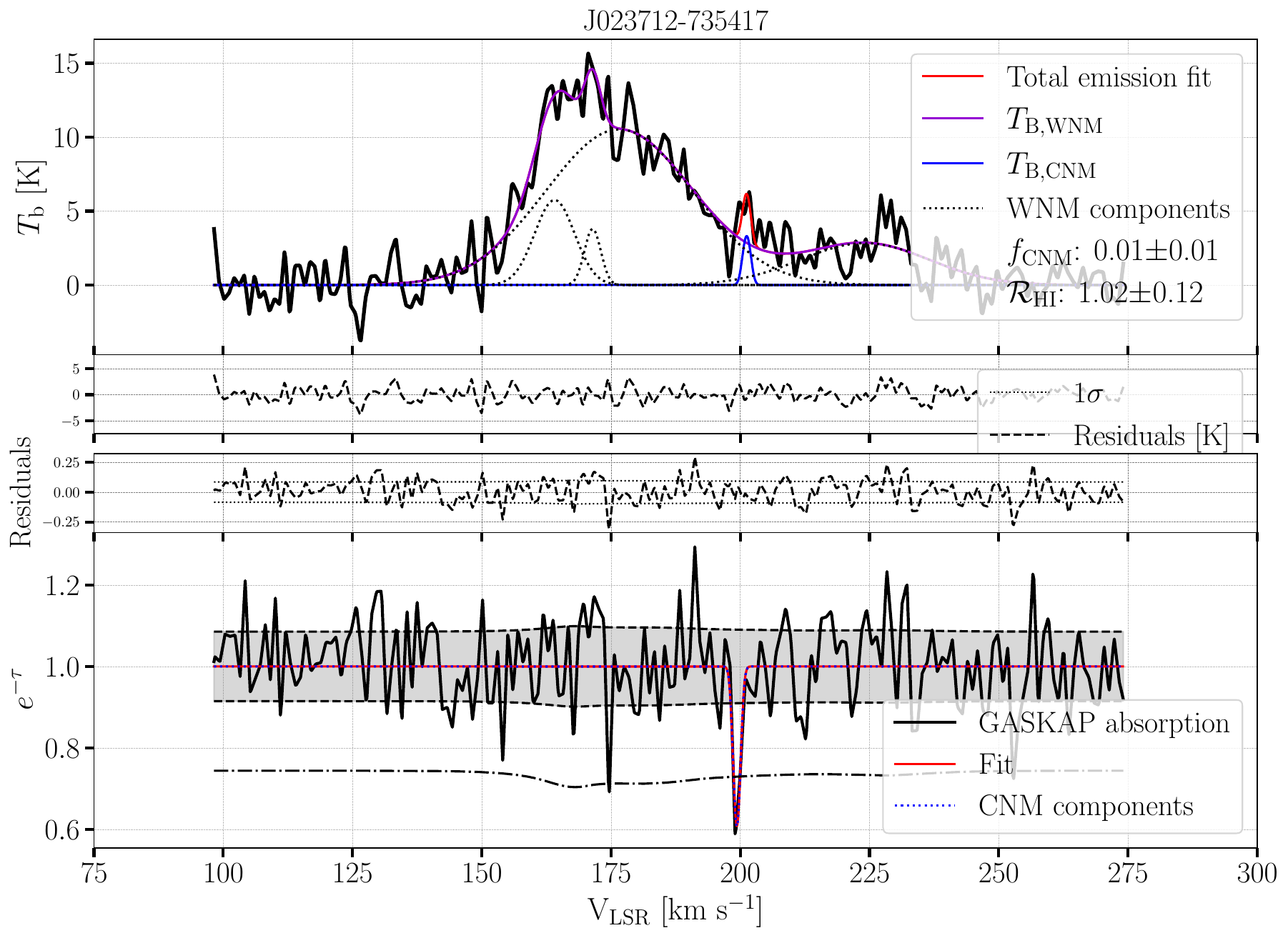}
  \includegraphics[width = 3.5in]{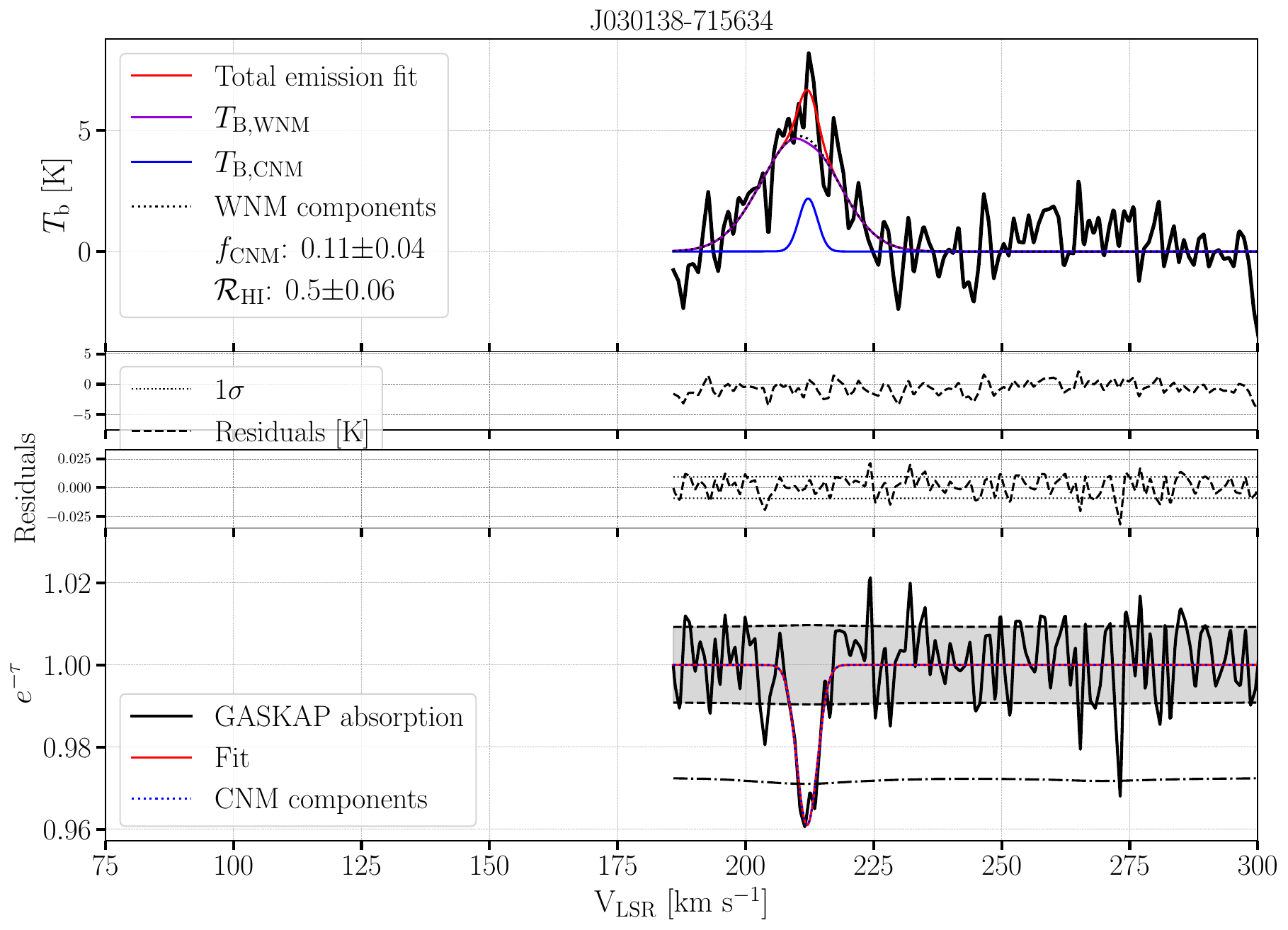}
  \includegraphics[width = 3.5in]{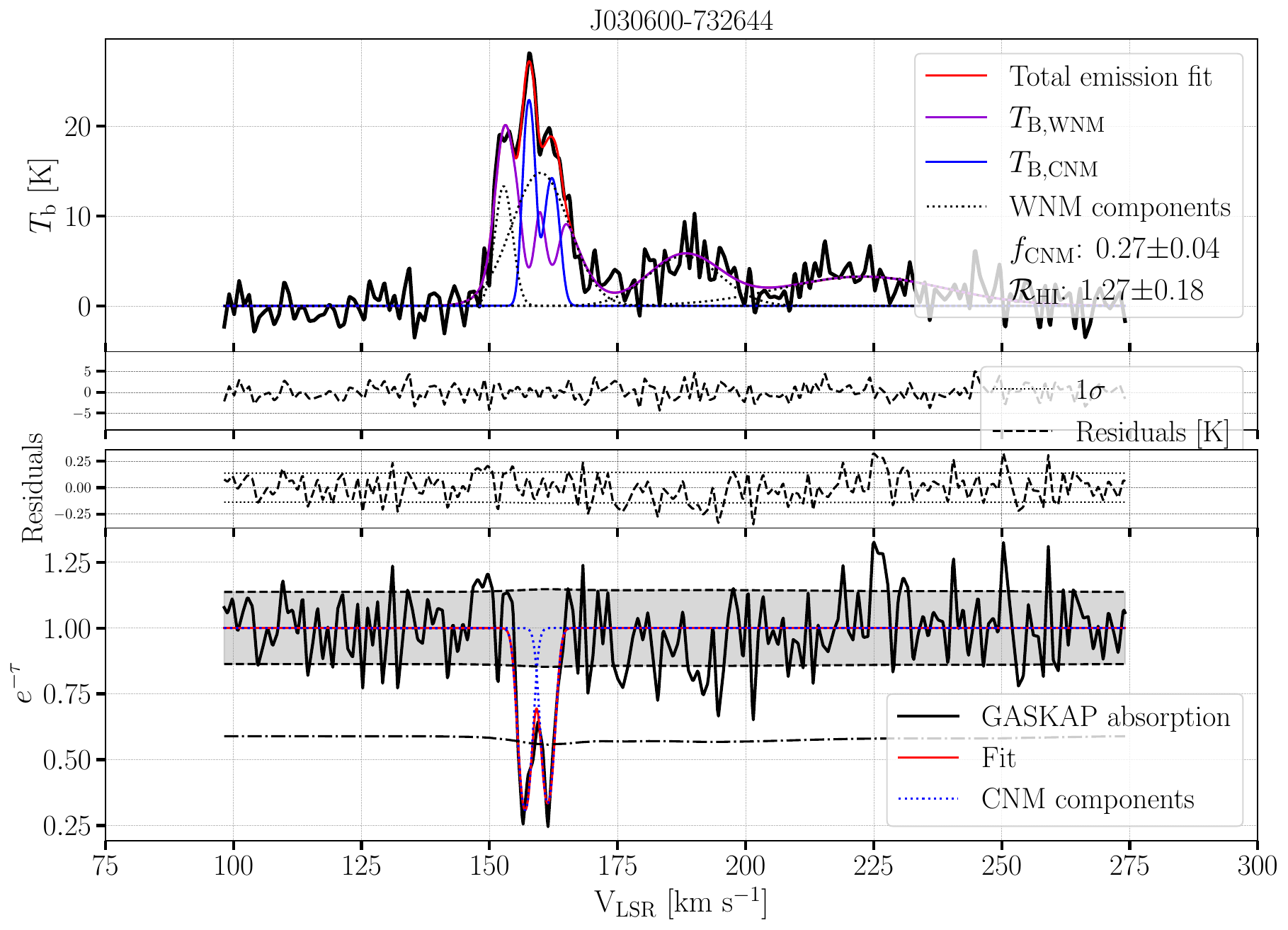}
  \includegraphics[width = 3.5in]{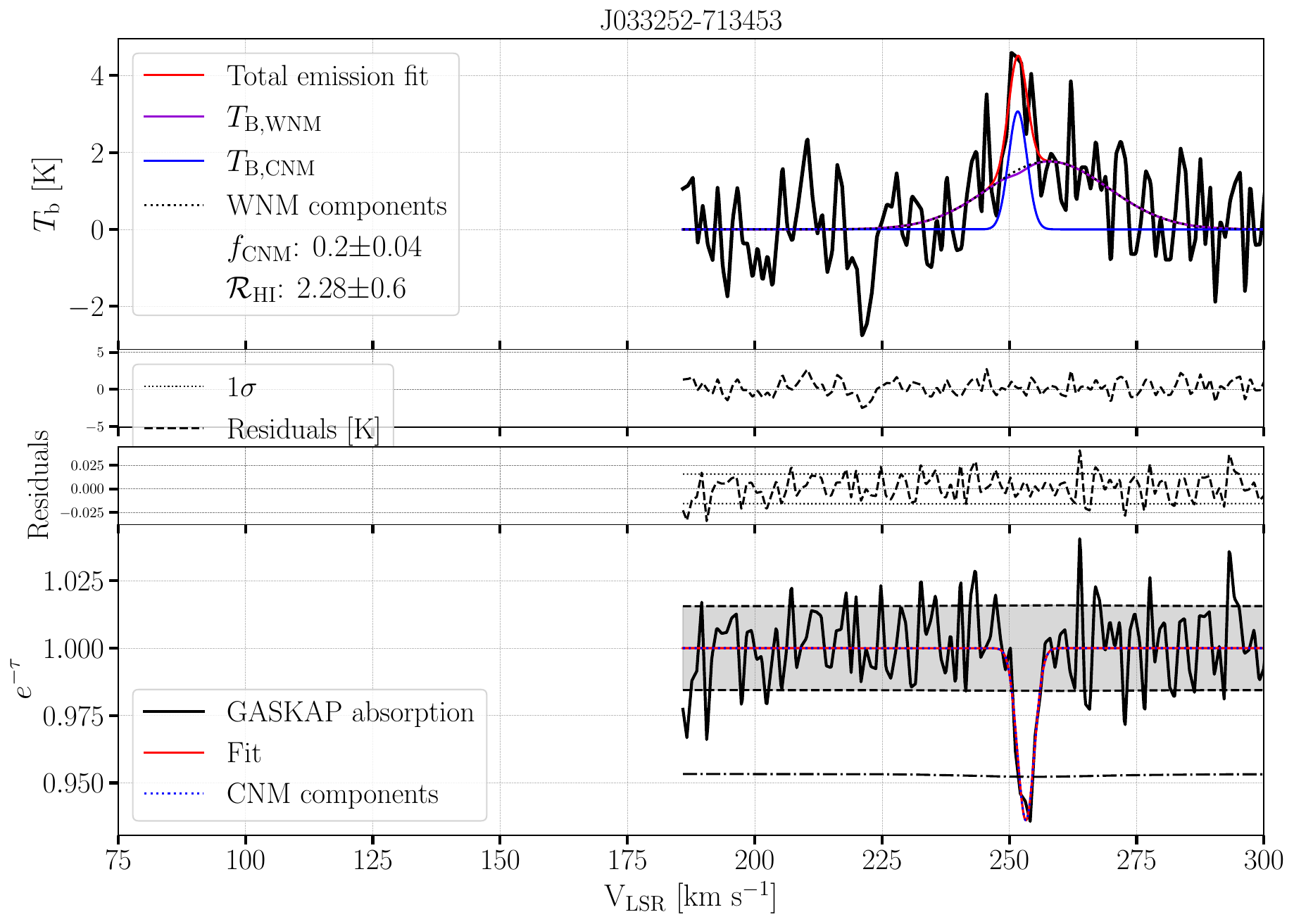}
  \caption{Gaussian decomposition of emission and absorption spectra for Magellanic Bridge sources (cont.)}
      
\end{figure*}

\bsp	
\label{lastpage}
\end{document}